\documentclass[aps,prd,superscriptaddress,onecolumn]{revtex4-2}
\usepackage{blindtext}

\usepackage{mathtools,mathalfa,amssymb,amsmath,amsfonts,amsthm,bm}
\usepackage[T1]{fontenc}
\usepackage{tipa}
\usepackage{wasysym}
\usepackage{appendix}
\usepackage{calrsfs}
\usepackage{tikz}
\usepackage[font=small,
   justification=justified,
   format=plain]{caption}
\usepackage{aas_macros}
\usetikzlibrary {decorations.pathmorphing}
\usepackage{orcidlink}

\usepackage{soul}

\usepackage{hyperref}
\hypersetup{
    colorlinks = true,
    citecolor  = blue,
    linkcolor = blue,
    urlcolor  = blue,
    linkbordercolor = white
}
       
\usepackage{framed}
\usepackage{color}
\usepackage[mathscr]{euscript}

\newcounter{definition}
\definecolor{wuppergreen}{RGB}{137, 186, 23}

\newcommand{\pullback}[1]{\hbox{\lower0.5ex\hbox{${}_{\leftarrow}$}}\kern-1.9ex{#1}}
\newcommand{\pullbacklong}[1]{\hbox{\lower0.85ex\hbox{${}_{\longleftarrow}$}}\kern-3.0ex{#1}}
\newcommand{\pullbackllong}[1]{\hbox{\lower0.85ex\hbox{${}_{\longleftarrow\!\!-\!\!-\!\!-\!\!-}$}}\kern-6.4ex{#1}}

\usepackage{graphicx}
\usepackage{subcaption}
\graphicspath{{./figures/}}

\begin{document}
\title{Efficient evaluation of real-time path integrals}

\author{Job Feldbrugge
\orcidlink{0000-0003-2414-8707}}
\email{job.feldbrugge@ed.ac.uk}
\affiliation{Higgs Centre for Theoretical Physics, University of Edinburgh, James Clerk Maxwell Building, Edinburgh EH9 3FD, UK}

\author{Joshua Y. L. Jones
\orcidlink{0000-0002-7854-0096}}
\email{jjones@stp.dias.ie}
\affiliation{School of Theoretical Physics, Dublin Institute for Advanced Studies, 10 Burlington Road, Dublin 4, Ireland}
\affiliation{School of Mathematics, Trinity College, Dublin 2, Ireland}

\begin{abstract}
    The Feynman path integral has revolutionized modern approaches to quantum physics. Although the path integral formalism has proven very successful and spawned several approximation schemes, the direct evaluation of real-time path integrals is still extremely expensive and numerically delicate due to its high-dimensional and oscillatory nature. We propose an efficient method for the numerical evaluation of the real-time world-line path integral for theories where the potential is dominated by a quadratic at infinity. This is done by rewriting the high-dimensional oscillatory integral in terms of a series of low-dimensional oscillatory integrals, that we efficiently evaluate with Picard-Lefschetz theory or approximate with the eikonal approximation. Subsequently, these integrals are stitched together with a series of fast Fourier transformations to recover the lattice regularized Feynman path integral. Our method directly applies to problems in quantum mechanics, the word-line quantization of quantum field theory, and quantum gravity.
\end{abstract}

\maketitle

\section{Introduction}
The Feynman path integral is arguably the most elegant formulation of quantum physics, linking quantum evolution to classical physics through interference \cite{Feynman:2005, Feynman:1948, Feynman:1985}. The transition amplitude between two states is expressed as a sum over histories, integrating over all paths interpolating between the initial and the final state. Classical evolution then arises by the constructive interference of paths close to a classical path interpolating between an initial and a final state. While Feynman originally introduced the path integral to non-relativistic quantum mechanics, the approach has found applications in many areas of science, including statistics, polymer physics, and financial markets (see for example the textbooks \cite{Feynman:1965, Schulman:2012, Kleinert:2004}). In physics, it is now central to quantum field theory \cite{Zee:2003}, and string theory \cite{Polchinski:1998}, and forms the bases for many explorations in quantum gravity and cosmology \cite{Gibbons:1993}. Moreover, it is the foundation of lattice gauge theory and quantum chromodynamics \cite{Smit:2002}.

While the Feynman path integral has changed our perspective on the universe from a philosophical perspective, it does not tell us how to carry out the sum. The path integral is a highly oscillatory and infinite-dimensional integral that converges due to intricate cancellations. This makes the definition of the path integral delicate and its evaluation typically expensive. Consequently, there are at  present several methods to approximate the path integral for different quantum systems. Firstly, the integrals are often studied perturbatively using Feynman diagrams. Secondly, much attention has been focused on Euclidean path integrals, where time is rotated to imaginary values and the convergence of the Feynman path integral often improves. Finally, path integrals are often approximated with the Wentzel–Kramers–Brillouin (WKB) method, focusing on the contribution of a set of (complex) classical paths sometimes known as instantons. 

Each approximation has its shortcomings. The perturbative Feynman diagram method has proven extremely effective in quantum electrodynamics. However, its application to quantum chromodynamics and the standard model of particle physics is still an active field of research as the number of Feynman diagrams rapidly grows with the order of the perturbative calculation. Note that some problems are intractable with perturbative methods. The Euclidean (imaginary time) path integral is successfully used in many non-perturbative applications and forms the basis of lattice gauge theory. Yet, for some quantum systems, the rotation to imaginary time does not yield a convergent Euclidean path integral. This is known as the sign-problem in quantum field theory and the conformal factor problem in quantum gravity. In other instances, where one wishes to describe the evolution of a quantum system, rotating the Euclidean results back to real time can be problematic, as small corrections in the Euclidean amplitude may dominate the real-time amplitude. Finally, the WKB approximation can provide a good understanding of the nature of the path integral in the non-perturbative regime. Even so, the WKB approximation fails near caustics, sometimes known as turning points. More importantly, when describing quantum tunneling events, it is often difficult to, from first principles, determine which complex classical paths to include in the approximation (see for example \cite{Turok:2014,Feldbrugge:2023_Rosen_Morse}). 

Recently, Picard-Lefschetz theory was proposed as a new method to tame the oscillatory nature of the real-time path integral \cite{Witten:2010}. Rather than rotating time to imaginary values, as we do for the Euclidean path integral, we deform the integration domain into the complex plane.  Using Cauchy's integral theorem, the deformation improves the convergence properties while preserving the quantum amplitude. Picard-Lefschetz theory was first formally developed and applied to finite-dimensional integrals, see for example \cite{Arnold:2012}. Feldbrugge et al.\ \cite{Feldbrugge:2023} turned this framework into a new numerical integration method suitable for families of highly oscillatory low-dimensional integrals. This integration method was specially constructed to evaluate astrophysical interference patterns in radio astronomy \cite{Feldbrugge:2020, Feldbrugge:2023_Multiplane, Jow:2023}. Remarkably, Picard-Lefschetz theory was then also combined with Hamiltonian Monte-Carlo techniques for the numerical evaluation of high-dimensional oscillatory integrals, capable of addressing the notorious “sign problem” in some quantum field theories \cite{Cristoforetti:2012, Cristoforetti:2013, Fujii:2013}. This led to \textit{the generalized Lefschetz thimble method} \cite{Alexandru:2016}, which was subsequently optimized in several fashions \cite{Alexandru:2017, Fukuma:2017, Fukuma:2019, Fukuma:2021, Fujisawa:2022}. These Monte Carlo Lefschetz methods elegantly combine the WKB instantons with integration methods typically employed in Euclidean lattice field theory, and have enabled the direct numerical evaluation of real-time path integrals for general potentials. For a recent study of quantum tunnelling focusing on the relevance of complex saddle points see for example \cite{Nishimura:2023, Nishimura:2024}. For the first application of Picard Lefschetz theory to quantum cosmology see \cite{Feldbrugge:2017}. For a recent application of the generalized Lefschetz thimble method to quantum cosmology see \cite{Chou:2024}. More formally, Picard-Lefschetz theory can likely be used to construct a continuum-time definition of the path integral. For a discussion of the measure in real-time path integrals see \cite{Feldbrugge:2023_Existence}.

In this paper, we propose an alternative efficient numerical method for the evaluation of real-time path integrals in quantum mechanics and in the word-line quantization of relativistic quantum theory, inspired by the new multi-plane lens integration technique \cite{Feldbrugge:2023_Multiplane}. For theories where the potential is dominated by a quadratic at infinity, we rewrite the time discretized path integral in terms of a set of low-dimensional oscillatory integrals that are subsequently stitched together using a series of fast Fourier transformations. This turns the evaluation of a $d\times (N-1)$-dimensional integral, with the dimension of configuration space $d$ and the number of discretization steps $N-1$, into the approximation/evolution of $N-1$ $d$-dimensional integrals and $2(N-1)$ $d$-dimensional fast Fourier transformations. The procedure is exact, robust and runs significantly faster than the generalized Lefschetz thimble method when exploring real-time path integrals as a function of the initial position, final position and propagation time. Notably, the computational cost scales linearly with the number of discretization steps and is trivially parallelizable. Extensions to real-time path integrals in quantum field theory and multi-particle path integrals will be investigated in future projects.

In section \ref{sec:Feynman}, we introduce the Feynman path integral in quantum mechanics. In section \ref{sec:stitching}, we introduce the stitching method which is central to this paper. In section \ref{sec:QM} we apply this method to the quantum mechanical path integral of a Rosen-Morse, smooth step, and double-well potential. In section \ref{sec:relativistic}, we discuss the application to the world-line quantization of relativistic quantum systems. The implementation, convergence rate and the general performance of the numerical integrator are discussed in section \ref{sec:performance}. Finally, in section \ref{sec:conclusion}, we discuss the results and remark on future applications.

%%%%%%%%%%%%%%%%%%%%%%%%%%%%%%%
\section{The Feynman path integral}\label{sec:Feynman}
In quantum mechanics, the evolution of a particle in a potential $V$ (for simplicity assumed to be time independent) is governed by the time-dependent Schrödinger equation
\begin{align}
    i \frac{\partial \psi_t(\bm{x})}{\partial t} = \left[-\frac{\hbar^2}{2m} \nabla^2 + V(\bm{x})\right]\psi_t(\bm{x})\,,
\end{align}
acting on the wave function $\psi$, with the initial condition 
\begin{align}
    \psi_{t=0}(\bm{x}) = \psi_0(\bm{x})\,.
\end{align}
The solution of the Schrödinger equation can be written as a convolution,
\begin{align}
    \psi_t(\bm{x}_1) = \int G(\bm{x}_1,\bm{x}_0, t)\psi_0(\bm{x}_0)\mathrm{d}\bm{x}_0\,,
\end{align}
of the initial state with the propagator $G$, where the propagator is a Green's function of the time-dependent Schrödinger equation. As such, it is defined by the differential equation
\begin{align}
    \left[i \hbar \frac{\partial}{\partial t} - \hat{H}\right] G(\bm{x}_1,\bm{x}_0,t) = i \hbar \delta(t)\,,
\end{align}
with the Hamiltonian operator
\begin{align}
    \hat{H} =  -\frac{\hbar^2}{2m} \nabla^2 + V(\bm{x})\,,
\end{align}
acting on either the initial position $\bm{x}_0$ or final position $\bm{x}_1$ (yielding two conditions), and the coincidence limit 
\begin{align}
    \lim_{t\to 0} G(\bm{x}_1,\bm{x}_0,t) = \delta(\bm{x}_0-\bm{x}_1)\,,
\end{align} 
with $\delta$ the Dirac delta function. 

The propagator satisfies the composition law 
\begin{align}
    G(\bm{x}_2,\bm{x}_0,T_1+T_2) = \int G(\bm{x}_2,\bm{x}_1,T_2) G(\bm{x}_1,\bm{x}_0,T_1)\mathrm{d}\bm{x}_1\,,\label{eq:composition}
\end{align}
that is to say, the amplitude for a particle to move from $\bm{x}_0$ to $\bm{x}_2$ in time $T_1+T_2$ coincides with the amplitude for a particle propagating from $\bm{x}_0$ to $\bm{x}_1$ in time $T_1$, and then from $\bm{x}_1$ to $\bm{x}_2$ in time $T_2$, summed over the intermediate position $\bm{x}_1$. This means the amplitude is the sum over all possible histories from $\bm{x}_0$ at time $t=0$ to $\bm{x}_2$ at time $t=T_1+T_2$ via any $\bm{x}_1$ at the intermediate time $t=T_1$. According to Feynman \cite{Feynman:1948,Feynman:1965}, the propagator can more generally be interpreted as a sum over histories, known as the Feynman path integral,
\begin{align}
    G(\bm{x}_1,\bm{x}_0,T) = \int_{\bm{x}(0)=\bm{x}_0}^{\bm{x}(T)=\bm{x}_1} e^{i S[\bm{x}]/\hbar}\mathcal{D}\bm{x}\,,
\end{align}
with the action of a non-relativistic particle 
\begin{align}
    S[\bm{x}] = \int_0^T \left[\frac{1}{2}m \dot{\bm{x}}^2 - V(\bm{x})\right]\mathrm{d}t\,.
\end{align}
Unfortunately, although highly suggestive, the preceding expression for the path integral is undefined as it stands (see the textbook \cite{Klauder:2010}). For example, the putative translation invariant measure $\mathcal{D}x$ does not exist, and the integral is at best conditionally convergent \footnote{When an integral diverges absolutely, $\int|f(x)|\mathrm{d}x = \infty$, the convergence of the integral $\int f(x) \mathrm{d}x$ subtly depends on the regularization of the integral.}. 

To overcome these basic problems, Feynman adopted a \textit{lattice regularization} as a procedure to yield well-defined integrals. Let's partition the time interval $[0,T]$ into $N$ equal time slices $t_n=a n$ for $n=0,1,\dots,N$ with $a=T/N$. Denote the position at $t_n$ by $\bm{q}_n = \bm{x}(t_n)$ with the boundary conditions $\bm{x}_0 = \bm{q}_0$ and $\bm{x}_1=\bm{q}_N$. Using the midpoint approximation, the discretized action between two time slices assumes the form
\begin{align}
    S_n = \frac{m}{2a}\left(\bm{q}_n - \bm{q}_{n+1}\right)^2 - \frac{a}{2} V(\bm{q}_n)- \frac{a}{2} V(\bm{q}_{n+1})\,.
\end{align}
The total discretized action
\begin{align}
    \sum_{n=0}^{N-1} S_n
    = \frac{m}{2a}\left[(\bm{x}_0-\bm{q}_1)^2 + \dots + (\bm{q}_{N-1} - \bm{x}_1)^2\right] - \frac{a}{2}\left[V(\bm{x}_0) + V(\bm{x}_1)\right] - a\left[V(\bm{q}_1) + \dots +V( \bm{q}_{N-1})\right]\,\label{eq:action}
\end{align}
can be used to define the discretized propagator
\begin{align}
    G_N(\bm{x}_1,\bm{x}_0,T) &= c 
    \int_{\mathbb{R}^{d(N-1)}} e^{ i\sum_{n=0}^{N-1}S_n/\hbar} \prod_{n=1}^{N-1} \left(c \mathrm{d} \bm{q}_n\right)\\
    &= c e^{-\frac{i a}{2 \hbar}\left[V(\bm{x}_0) + V(\bm{x}_1)\right]}
    \int_{\mathbb{R}^{d(N-1)}} e^{ \frac{i m}{2a \hbar}\left[(\bm{x}_0-\bm{q}_1)^2 + \dots + (\bm{q}_{N-1} - \bm{x}_1)^2\right]- \frac{i a}{\hbar}\left[V(\bm{q}_1) + \dots +V( \bm{q}_{N-1})\right]} \prod_{n=1}^{N-1} \left(c \mathrm{d} \bm{q}_n\right)\,,
    \label{eq:lattice}
\end{align}
evaluated over the intermediate positions $\bm{q}_1,\dots, \bm{q}_{N-1}$, with the normalization $c = \left(\frac{m}{2 \pi i \hbar a}\right)^{d/2}$, and the dimension of configuration space $d$. The continuum path integral is recovered in the limit
\begin{align}
    G(\bm{x}_1,\bm{x}_0,T) &= \lim_{N\to \infty} G_N(\bm{x}_1,\bm{x}_0,T)\,.
\end{align}

In this paper, we focus exclusively on the lattice regularization of the path integral. For alternative continuum-time regularization schemes we refer to \cite{Klauder:2003, Feldbrugge:2023_Existence}.

%%%%%%%%%%%%%%%%%%%%%%%%%%%%%%%
\section{The stitching method}\label{sec:stitching}
The lattice regularized path integral \eqref{eq:lattice} is a $d(N-1)$-dimensional integral whose integration measure is a well-defined Lebesgue product measure. Nonetheless, the integral is still not completely unambiguous, as the integrand does not converge absolutely. The integral converges conditionally due to delicate cancellations and its definition depends on the choice of regulator, as the dominated convergence theorem does not apply. In this paper, we will assume a smooth analytic regularization scheme which is equivalent to an analytic deformation of the integrand or a smooth deformation of the integration domain. See Feldbrugge and Turok \cite{Feldbrugge:2023_Existence} for a detailed discussion on smooth regulators of conditionally convergent integrals. This regularization scheme is also used in the generalized Lefschetz thimble method \cite{Alexandru:2016}.

Even after obtaining a rigorous definition, the evaluation of the discretized path integral has so far been difficult for most potentials. The integral is high-dimensional, and away from its saddle points, the integrand oscillates wildly. We here propose an efficient method for the evaluation of the discretized real-time path integral for theories where the potential is dominated by a quadratic at infinity. After approximating/evaluating $N-1$ families of $d$-dimensional integrals, we stitch them together with $2(N-1)$ fast Fourier transformations to recover the $d(N-1)$-dimensional discretized path integral as a function of the propagation time and final position.

%%%%%%%%%%%%%%%%%%%%%%%%%%%%%%%
\subsection{Chain of linked integrals}
First note that the discretized action \eqref{eq:action} has a simple structure. Neighbouring integration variables are coupled through the discretized kinetic terms, \textit{i.e.}, the integration variable $\bm{q}_n$ only couples to its direct neighbours $\bm{q}_{n-1}$ and $\bm{q}_{n+1}$ through the terms $(\bm{q}_{n-1}-\bm{q}_n)^2$ and $(\bm{q}_{n}-\bm{q}_{n+1})^2$. Consequently, we can phrase the discretized path integral in an iterative form. Starting with
\begin{align}
    I_1(\bm{q}_1) &=  e^{\frac{i m}{2 a \hbar}(\bm{x}_0-\bm{q}_1)^2 } 
\end{align}
and iterating with the rule 
\begin{align}
    I_{n+1}(\bm{q}_{n+1}) &= c \int I_{n}(\bm{q}_n)e^{\frac{i m}{2 a \hbar}(\bm{q}_n - \bm{q}_{n+1})^2 - \frac{i a}{\hbar} V(\bm{q}_n)} \mathrm{d}\bm{q}_n\,,
\end{align}
we obtain the integral $I_{N}(\bm{q}_N)$, yielding the discretized path integral
\begin{align}
    G_N(\bm{x}_1,\bm{x}_0,T) = c e^{-\frac{i a}{2 \hbar}\left[V(\bm{x}_0) + V(\bm{x}_1)\right]}I_{N}(\bm{x}_1)\,.
\end{align} 
More generally, the intermediate integral $I_n$ yields an approximation of the propagator $G(\bm{x}_1,\bm{x}_0,an)$, given by
\begin{align}
    G_n(\bm{x}_1,\bm{x}_0,an) = c e^{-\frac{i a}{2 \hbar}\left[V(\bm{x}_0) + V(\bm{x}_1)\right]}I_{n}(\bm{x}_1)\,,
\end{align} 
for $n=1,\dots,N$. This iterative scheme is justified by the composition law \eqref{eq:composition}, slowly propagating from the initial time to the total propagation time $T$. 

%%%%%%%%%%%%%%%%%%%%%%%%%%%%%%%
\subsection{Asymptotics of the intermediate integrals}
So far, we have merely rewritten the discretized path integral into an iterative form. Next, consider the first integral, 
\begin{align}
    I_2(\bm{q}_2)
    = c \int  e^{\frac{i m}{2 a \hbar}\left((\bm{x}_0-\bm{q}_1)^2 + (\bm{q}_1 - \bm{q}_2)^2\right) - \frac{i a}{\hbar} V(\bm{q}_1)} \mathrm{d}\bm{q}_1\,, \label{eq:first_integral}
\end{align}
which we can numerically evaluate as a function of $\bm{q}_2$ using, for example, the Picard-Lefschetz methods described in \cite{Feldbrugge:2023}. Alternatively, we can perform a numerical integral over a rough deformation of the integration domain, or use the eikonal approximation (see section \ref{sec:Js}). The saddle point of the exponent with respect to the variable $\bm{q}_1$,
\begin{align}
    \frac{m}{a}( 2\bm{q}_1 - \bm{x}_0 - \bm{q}_2) - a \nabla V (\bm{q}_1) = 0\,,
\end{align}
yields upon solving for $\bm{q}_2$ the map 
\begin{align}
    \bm{\xi}_2(\bm{q}_1) &= -\bm{x}_0 + 2 \bm{q}_1 - \frac{a^2}{m} \nabla V(\bm{q}_1)\,,
\end{align}
associating each point $\bm{q}_1$ at time $t=a$ to a point $\bm{q}_2$ at time $t=2a$. The eikonal approximation of \eqref{eq:first_integral} -- including only real saddle points -- assumes the form
\begin{align}
    I_2(\bm{q}_2) &\approx 
    \sum_{\bm{q}_1 \in \bm{\xi}_2^{-1}(\bm{q}_2)} 
    \frac{
        e^{\frac{i m}{2 a \hbar}\left((\bm{x}_0-\bm{q}_1)^2 + (\bm{q}_1 - \bm{q}_2)^2\right) - \frac{i a}{\hbar} V(\bm{q}_1)}}
        {\sqrt{\det \nabla \bm{\xi}_2(\bm{q}_1)}}\\
    &= 
    \sum_{\bm{q}_1 \in \bm{\xi}_2^{-1}(\bm{q}_2)} 
    \frac{
        e^{\frac{i m}{2 a \hbar}\left((\bm{x}_0-\bm{q}_1)^2 + (\bm{q}_1 - \bm{q}_2)^2\right) - \frac{i a}{\hbar} V(\bm{q}_1)}}
        {\sqrt{\det \left(2I -\frac{a^2}{m} \mathcal{H} V(\bm{q}_1)\right)}}\,.
\end{align}
The integral $I_2$ can include complex behaviour with caustics and intricate interference patterns, particularly for large time steps $a$. However, assuming that the gradient of the potential $\lVert \nabla V(\bm{q}_1) \rVert$ is dominated by the linear term $2 \bm{q}_1$ as $\lVert\bm{q}_1\rVert \to \infty$ (\textit{i.e.} we have $\mathcal{H}V(\bm{q}_{1}) \ll 2 \text{ as } \lVert\bm{q}_{1}\rVert\to \infty$, as we will have in the examples shown later), the integral approaches the asymptotic
\begin{align}
    I_2(\bm{q}_2) \sim 
    \bar{J}_2(\bm{q}_2) = 
    \frac{1}{2^{d/2}} e^{\frac{i m}{2 a \hbar}\frac{(\bm{x}_0 - \bm{q}_2)^2}{2} - \frac{i a}{\hbar} V\left(\frac{\bm{x}_0 + \bm{q}_2}{2}\right)}
\end{align}
for large $\lVert \bm{q}_2 \rVert$. In cases where the potential is asymptotically quadratic, the eikonal approximation similarly provides a appropriate $\bar{J}$s. Writing the integral \eqref{eq:first_integral}, as a residual with respect to this asymptotic, 
\begin{align}
    I_2(\bm{q}_2) = \bar{J}_2(\bm{q}_2) + \delta I_2(\bm{q}_2)\,,
\end{align}
the second integral in the iterative series consists of two terms
\begin{align}
    I_{3}(\bm{q}_{3}) &= 
    c \int \bar{J}_{2}(\bm{q}_2)e^{\frac{i m}{2 a \hbar}(\bm{q}_2 - \bm{q}_{3})^2 - \frac{i a}{\hbar} V(\bm{q}_2)} \mathrm{d}\bm{q}_2
    +
    c \int \delta I_{2}(\bm{q}_2)e^{\frac{i m}{2 a \hbar}(\bm{q}_2 - \bm{q}_{3})^2 - \frac{i a}{\hbar} V(\bm{q}_2)} \mathrm{d}\bm{q}_2\,.
\end{align}
The first term, 
\begin{align}
    J_3(\bm{q}_3) &= 
    c \int \bar{J}_{2}(\bm{q}_2)e^{\frac{i m}{2 a \hbar}(\bm{q}_2 - \bm{q}_{3})^2 - \frac{i a}{\hbar} V(\bm{q}_2)} \mathrm{d}\bm{q}_2\\
    &= 
    \frac{c}{2^{d/2}}\int e^{\frac{i m}{2 a \hbar}\left(\frac{1}{2}(\bm{x}_0 - \bm{q}_2)^2 + (\bm{q}_2 - \bm{q}_{3})^2\right) - \frac{i a}{\hbar}\left(V\left(\frac{\bm{q}_2+\bm{x}_0}{2}\right) +  V(\bm{q}_2)\right)} \mathrm{d}\bm{q}_2\,,
\end{align}
resembles equation \eqref{eq:first_integral} and can be evaluated numerically or approximated (see section \ref{sec:Js}). The second integral converges absolutely as $\delta I_2$ decays for large $\lVert \bm{q}_2 \rVert$ by construction. Moreover, as the integral assumes the form of a convolution, we efficiently evaluate the integral with two fast Fourier transformations, 
\begin{align}
    c \int \left(\delta I_{2}(\bm{q}_2)e^{ - \frac{i a}{\hbar} V(\bm{q}_2)}\right)e^{\frac{i m}{2 a \hbar}(\bm{q}_2 - \bm{q}_{3})^2}  \mathrm{d}\bm{q}_2
    =  \mathcal{F}^{-1}\left[ \mathcal{F}\left(\delta I_2(\bm{q}_2) e^{ - \frac{i a}{\hbar} V(\bm{q}_2)}\right) e^{-\frac{i a \hbar \bm{k}^2}{2m}} \right]\,,
\end{align}
with the fast Fourier transform $\mathcal{F}$ and its inverse $\mathcal{F}^{-1}$, as the Fourier transform of a Gaussian kernel is Gaussian
\begin{align}
    c \int e^\frac{i m\bm{q}^2}{2 a \hbar} e^{i \bm{q} \cdot \bm{k}}\mathrm{d}\bm{q} = e^{-\frac{i a \hbar\bm{k}^2}{2m}}\,.
\end{align}
Note that we cannot apply the fast Fourier trick to the original integral, as $I_2(\bm{q}_2)$ oscillates rapidly and does not decay for large $\lVert\bm{q}_2\rVert$. The integral $I_3(\bm{q}_3)$ is then the sum of these two contributions.

Next, consider the $I_4$ integral. The saddle point of the $J_3$ integral leads to the map 
\begin{align}
    \bm{\xi}_3(\bm{q_2}) = \frac{-\bm{x}_0 + 3 \bm{q}_2}{2} - \frac{a^2}{m} \left[\frac{1}{2}\nabla V\left(\frac{\bm{x}_0 + \bm{q}_2}{2}\right) +  \nabla V(\bm{q}_2)\right]\,.
\end{align}
For large $\lVert\bm{q}_2\rVert$, the associated eikonal approximation approaches the asymptotic
\begin{align}
    J_3(\bm{q}_3) \sim 
    \bar{J}_3(\bm{q}_3) = 
    \frac{1}{3^{d/2}} e^{\frac{i m}{2 a \hbar}\frac{(\bm{x}_0 - \bm{q}_3)^2}{3} - \frac{i a}{\hbar} \left[V\left(\frac{2\bm{x}_0 + \bm{q}_3}{3}\right) + V\left(\frac{\bm{x}_0 + 2\bm{q}_3}{3}\right)\right]}\,.
\end{align}
Writing $I_3(\bm{q}_3) = \bar{J}_3(\bm{q}_3) + \delta I_3(\bm{q}_3)$, we can again write the integral $I_4$ as the sum of a $d$-dimensional integral involving $\bar{J}_3$ and two fast Fourier transformations involving $\delta I_3$.

%%%%%%%%%%%%%%%%%%%%%%%%%%%%%%%
\subsection{The stitching method}
Iterating this procedure until we reach $n=N$, we write the integral $J_n(\bm{q}_n)$ 
\begin{align}
    J_{n}(\bm{q}_{n}) &= c \int \bar{J}_{n-1}(\bm{q}_{n-1})e^{\frac{i m}{2 a \hbar}(\bm{q}_{n-1} - \bm{q}_{n})^2 - \frac{i a}{\hbar} V(\bm{q}_{n-1})} \mathrm{d}\bm{q}_{n-1}\\
    &= \frac{c}{(n-1)^{d/2}} \int e^{\frac{i m}{2 a \hbar}\left(\frac{(\bm{x}_0-\bm{q}_{n-1})^2}{n-1} + (\bm{q}_{n-1} - \bm{q}_{n})^2\right) - \frac{i a}{\hbar} \sum_{k=1}^{n-1}V(\bar{\bm{q}}_{n-1}^k)} \mathrm{d}\bm{q}_{n-1}\,
\end{align}
in terms of the asymptotic $\bar{J}_{n-1}$ of the previous $J_{n-1}$ function,
\begin{align}
    \bar{J}_{n}(\bm{q}_n) = \frac{1}{n^{d/2}} e^{\frac{im}{2 a \hbar}\frac{(\bm{x}_0 - \bm{q}_n)^2}{n}-\frac{i a }{\hbar} \sum_{k=1}^{n-1}V(\bar{\bm{q}}_n^k)}\,,
\end{align}
with the linear interpolation
\begin{align}
    \bar{\bm{q}}^{k}_n = \bm{x}_0 + \frac{k}{n}(\bm{q}_n - \bm{x}_0)\,.
\end{align}
The subscript in the linear interpolation $\bar{\bm{q}}_n^k$ emphasizes its dependence on the point $\bm{q}_n$.

\bigskip
\noindent To evaluate the discretized path integral we propose the following steps:
\begin{enumerate}
    \item First, approximate/evaluate the integrals $J_2, \dots, J_N$ on a regular lattice using either an explicit deformation of the integration contour or a saddle point approximation as discussed in section \ref{sec:Js}. Alternatively, we can use the numerical implementation of Picard-Lefschetz theory for families of low-dimensional integrals presented in \cite{Feldbrugge:2023}.
    \item Next, stitch the $J_n$'s together to evaluate the integrals $I_2,\dots, I_N$.  
    Starting with 
    \begin{align}
        I_2(\bm{q}_2) = J_2(\bm{q}_2)\,,
    \end{align}
    we evaluate the residual $\delta I_2$ using the rule
    \begin{align}
        \delta I_n(\bm{q}_n) = I_n(\bm{q}_n) - \bar{J}_n(\bm{q}_n)\,,
    \end{align}
    and find the next step $I_3$ using the iterative process
    \begin{align}
        I_{n+1}(\bm{q}_{n+1})=J_{n+1}(\bm{q}_{n+1}) +  \mathcal{F}^{-1}\left[ \mathcal{F}\left(\delta I_{n}(\bm{q}_n) e^{ - \frac{i a}{\hbar} V(\bm{q}_n)}\right) e^{-\frac{i a \hbar \bm{k}^2}{2m}} \right]\,,
    \end{align}
    with the fast Fourier transformation $\mathcal{F}$ and its inverse $\mathcal{F}^{-1}$.
    We repeat this process to obtain the integrals $I_2,\dots,I_N$
    \item Upon normalizing $I_n$
    \begin{align}
        G_n(\bm{x}_1,\bm{x}_0,an) = c e^{-\frac{i a}{2 \hbar}\left[V(\bm{x}_0) + V(\bm{x}_1)\right]}I_{n}(\bm{x}_1)\,,
    \end{align}
    we obtain the discretized path integral for $n=1,\dots, N$. 
\end{enumerate} 
This process simultaneously yields the discretized path integral $G_n$ for a lattice of propagation times and final positions for a fixed initial position $\bm{x}_0$. By repeating this procedure for a lattice of initial positions $\bm{x}_0$, we evaluate the discretized propagator $G_n$. 

%%%%%%%%%%%%%%%%%%%%%%%%%%%%%%%
\subsection{Approximating/evaluating the $J_n$ integral}\label{sec:Js}
The previous section introduced the stitching method, reducing the evaluation of the discretized path integral $G_N(\bm{x}_1,\bm{x}_0,T)$ to the approximation/evaluation of the integrals $J_n$. We now focus on the efficient evaluation and approximation of the $J_n$ integrals.

Firstly, by completing the square, we can rewrite the $J_n$ integral as
\begin{align}
    J_{n}(\bm{q}_{n}) &= \frac{c}{(n-1)^{d/2}} e^{\frac{i m}{2 a \hbar}\frac{(\bm{q}_{n} - \bm{x}_0)^2}{n}}
    \int e^{\frac{i m}{2 a \hbar}\frac{n}{n-1} \left(\bm{q}_{n-1} - \bm{q}_s\right)^2 - \frac{i a}{\hbar} \sum_{k=1}^{n-1}V(\bar{\bm{q}}_{n-1}^k)} \mathrm{d}\bm{q}_{n-1}\\
    &= \frac{c}{(n-1)^{d/2}} e^{\frac{i m N}{2 T \hbar}\frac{(\bm{q}_{n} - \bm{x}_0)^2}{n}}
    \int e^{\frac{i m N}{ T \hbar}\frac{n}{n-1} \left[\frac{\left(\bm{q}_{n-1} - \bm{q}_s\right)^2}{2} - \frac{ T^2}{ m N^2} \frac{n-1}{n} \sum_{k=1}^{n-1}V(\bar{\bm{q}}_{n-1}^k)\right]} \mathrm{d}\bm{q}_{n-1}\,, \label{eq:Jn_form}
\end{align}
with the saddle point of the kinetic term
\begin{align}
    \bm{q}_s =  \frac{\bm{x}_0 + (n-1) \bm{q}_{n}}{n}\,.
\end{align}
This integral is usually less oscillatory than the original integral as the oscillations in $J_n$ are typically governed by the prefactor for large $\bm{q}_n$. When this is the case, estimating $J_n$ requires fewer approximations/evaluations.

%%%%%%%%%%%%%%%%%%%%%%%%%%%%%%%
\subsubsection{Numerical evaluation}
For the numerical evaluation, deform the integration domain by rotating the contour around the saddle point of the kinetic term $\bm{q}_s$ such that it is parametrized by
\begin{align}
    \bm{\gamma}_{n-1}(\bm{\lambda}) = \bm{q}_s + e^{i \bm{\theta}} \bm{\lambda}\,,
\end{align}
for $\bm{\lambda} \in \mathbb{R}^{d}$ and some angles $\bm{\theta} \in [0, 2\pi)^d$. Letting $\bm{q}_{n-1} = \bm{\gamma}_{n-1}(\bm{\lambda})$, the integral assumes the form 
\begin{align}
    J_{n}(\bm{q}_{n}) 
    = \frac{c}{(n-1)^{d/2}} e^{\frac{i m}{2 a \hbar}\frac{(\bm{q}_{n} - \bm{x}_0)^2}{n} + i (\theta_1 + \dots + \theta_d)}
    \int e^{\frac{i m}{2 a \hbar}\frac{n}{n-1} e^{2 i \theta} \bm{\lambda}^2 - \frac{i a}{\hbar} \sum_{k=1}^{n-1}V(\bar{\bm{q}}_{n-1}^k)}  \mathrm{d}\bm{\lambda}\,,
\end{align}
where $\bar{\bm{q}}_{n-1}^{m}$ depends on $\bm{\lambda}$ through $\bm{\gamma}_{n-1}(\bm{\lambda})$. 

\begin{figure}
    \centering
    \includegraphics{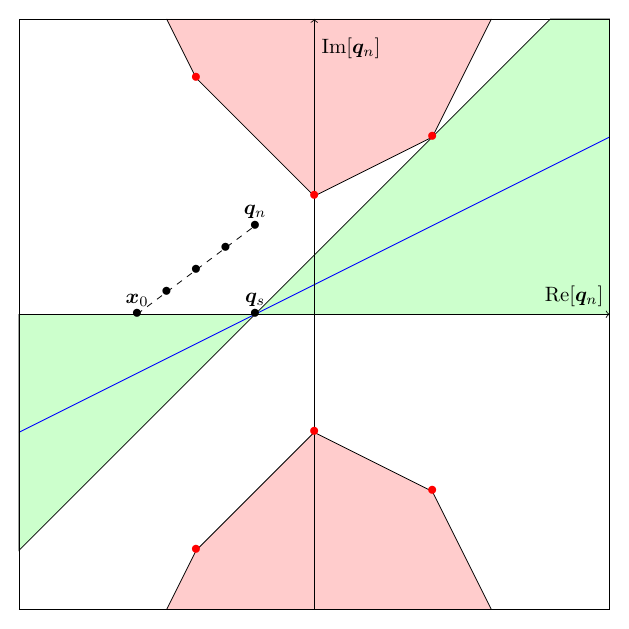}
    \caption{The proposed linear deformation of the integration domain in the complex $\bm{q}_n$ plane. The convex hull (the shaded red region) of the singularities of the potential $V$ (the red points) mark regions where the sum $\sum_m V(\bar{\bm{q}}_n^m)$ may become singular. This is illustrated by an example path $\bar{\bm{q}}_n$ linearly interpolating between $\bm{x}_0$ and $\bm{q}_n$. A linear deformation of the integration domain of the form $\bm{q}_s + e^{i \theta}\mathbb{R}^d$ (marked by the blue line) is guaranteed to preserve the integral value when it does not cross this convex hull of the singularities of the potential and resides in the green shaded region.}\label{fig:convex_hull}
\end{figure}

This integral is independent of $\bm{\theta}$ (for small angles) by Cauchy's integral theorem, but the nature of the integrand varies dramatically as a function of the angles. When it comes to the kinetic term, the optimal angle $\bm{\theta}$ would be $(\pi/4, \dots, \pi/4)$, as $i e^{2 i \pi/4}=-1$, rendering the oscillatory integral a convex $d$-dimensional Gaussian integral. However, while deforming the integration domain we should ensure that the contour does not pass through a singularity of the integrand. Thus by finding the singularities of $\sum V(\bar{\bm{q}})$ we can find a valid range of angles $\bm{\theta}$. We find a good deformation with two steps:
\begin{enumerate}
    \item First find the singularities of $V$ in the complex plane and identify the convex hulls of the singularities in the upper and lower complex half-planes. For a $\bm{q}_n$ outside these regions, the potential $\sum_m V(\bar{\bm{q}})$ is free of singularities (see fig.\ \ref{fig:convex_hull}).
    \item Next, deform the original integration domain into the complex plane $\mathbb{C}^{d}$ while making sure that the deformed contour does not enter either of the two convex hulls (see fig.\ \ref{fig:convex_hull}). This deformation suppresses the oscillations of the integrand for large $\bm{\lambda}$.
\end{enumerate}
The resulting integral can be efficiently numerically evaluated with standard integration techniques. Here we use the Gauss-Hermite quadrature method
\begin{align}
    \int_{-\infty}^\infty f(x) e^{-x^2}\mathrm{d}x \approx \sum_{i=1}^M\omega_i f(x_i)\,,
\end{align}
with weights $\omega_i$ and associated locations $x_i$ \cite{NIST:2011}. One can use more general/optimal deformations of the integration contour going around these singularities \cite{Feldbrugge:2023}, but these integration methods are often more computationally expensive.

%%%%%%%%%%%%%%%%%%%%%%%%%%%%%%%
\subsubsection{The eikonal approximation}
Alternatively, note how equation \eqref{eq:Jn_form} scales with the number of discretization steps $N$. As $N$ increases, the effective reduced Planck constant,
\begin{align}
    \hbar_{eff}  = \frac{ T \hbar}{m N} \frac{n-1}{n}\,,\label{eq:eff_Planck}
\end{align}
governing the length scale of the oscillations of the integrand decreases, and so the phase-variation
\begin{align}
    \frac{ T^2}{ m N^2} \frac{n-1}{n} \sum_{k=1}^{n-1}V(\bar{\bm{q}}_{n-1}^k)\,, \label{eq:eff_phase}
\end{align}
of the integral $J_n$ similarly decreases. Consequently, for large $N$, the integral $J_n$ is free of caustics and is increasingly well approximated with the eikonal approximation, including a single real saddle point.

The gradient of the exponent of the integrand with respect to $\bm{q}_{n-1}$,
\begin{align}
    {\frac{ m}{ a }\frac{n}{n-1} \left(\bm{q}_{n-1} - \bm{q}_s\right) -  a \sum_{k=1}^{n-1} \nabla_{\bm{q}_{n-1}}V(\bar{\bm{q}}_{n-1}^k)} = 0\,,
\end{align}
upon solving for $\bm{q}_s$ yields the map
\begin{align}
    \bm{\chi}_n(\bm{q}_{n-1}) 
    =\bm{q}_{n-1} -\frac{ a^2}{m }\frac{1}{n} \sum_{k=1}^{n-1} k (\nabla V)(\bar{\bm{q}}_{n-1}^k)\,.
\end{align}
The eikonal approximation of $J_n$ then assumes the form 
\begin{align} 
    J_{n}(\bm{q}_n) \approx \frac{1}{(n-1)^{d/2}} e^{\frac{im}{2 a \hbar} \frac{(\bm{q}_n-\bm{x}_0)^2}{n}}
    \sum_{\bm{q}_{n-1} \in \bm{\chi}_n^{-1}(\bm{q}_s)}\frac{e^{\frac{i m}{2 a \hbar}\frac{n}{n-1} \left(\bm{q}_{n-1} - \bm{q}_s\right)^2 - \frac{i a}{\hbar} \sum_{k=1}^{n-1}V(\bar{\bm{q}}_{n-1}^k)}}{\sqrt{\det \nabla \bm{\chi}_n(\bm{q}_{n-1})}}\,.
\end{align}
With an efficient implementation of the inverse $\bm{\chi}^{-1}$, the eikonal approximation can outperform the numerical evaluation of $J_n$. This speedup is particularly significant for problems with a multidimensional parameter space, $d>1$.

%%%%%%%%%%%%%%%%%%%%%%%%%%%%%%%
\section{Quantum mechanics} \label{sec:QM}
We now demonstrate the stitching method for three one-dimensional quantum mechanical systems, the fully solvable Rosen-Morse barrier potential, the smooth step potential, and the double-well potential.

%%%%%%%%%%%%%%%%%%%%%%%%%%%%%%%
\subsection{Rosen-Morse theory}\label{sec:Rosen-Morse}
The evolution of a non-relativistic particle in a Rosen-Morse potential
\begin{align}
    V(x) = \frac{V_0}{\cosh^2 x}\,,
\end{align}
is an ideal test case for the stitching method, as it is one of the few nontrivial exactly solvable quantum systems. The time-independent Schrödinger equation 
\begin{align}
    \hat{H} \phi_k(x) = \frac{\hbar^2 k^2}{2m} \phi_k(x)\,,
\end{align} 
is solved by the eigenstates 
\begin{align}
    \phi^+_k(x) &= \sqrt{\frac{k \sinh(\pi k)}{\cosh(2 \pi k )+ \cosh(2 \pi \nu)}}P_N^{ik}(\tanh x)\,,\\
    \phi^-_k(x) &= \sqrt{\frac{k \sinh(\pi k)}{\cosh(2 \pi k )+ \cosh(2 \pi \nu)}}P_N^{ik}(-\tanh x)\,, 
\end{align}
for positive $k$, with the associated Legendre function $P_{\sigma}^\mu(x)$ and the constant
\begin{align}
    N=-\frac{1}{2} + \frac{i}{2\hbar} \sqrt{8 m V_0-\hbar^2}\,.
\end{align}
For simplicity, we will always assume $ V_0 >\frac{\hbar^2}{8 m}$. We numerically evaluate the spectral representation of the real-time propagator 
\begin{align}
    G(x_1,x_0,T) = \Theta_H(T)\int_{0}^\infty \left[\phi^+_k(x_1)\phi^+_k(x_0)^*+\phi^-_k(x_1)\phi^-_k(x_0)^*\right] e^{-\frac{i \hbar k^2 T}{2m}}\mathrm{d}k\,,
\end{align}
by slightly deforming the integration domain $(0,\infty)$ into the complex plane. For a detailed study of the Rosen-Morse potential see \cite{Poschl:1933, Rosen:1932, Kleinert:1992, Grosche:1998, Grosche:1993, Kleinert:2004, Feldbrugge:2023_Rosen_Morse}. For a detailed discussion on the numerical evaluation of this integral see \cite{Feldbrugge:2023_Rosen_Morse}.

\begin{figure}
    \centering
    \begin{subfigure}[b]{0.49\textwidth}
        \centering
        \includegraphics[width=\textwidth]{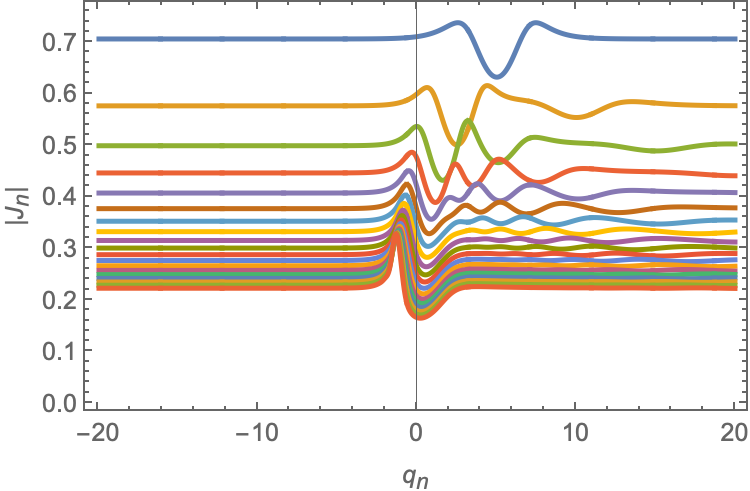}
        \caption{The modulus of the $J_n$ integral.}
    \end{subfigure}
    \hfill
    \begin{subfigure}[b]{0.49\textwidth}
        \centering
        \includegraphics[width=\textwidth]{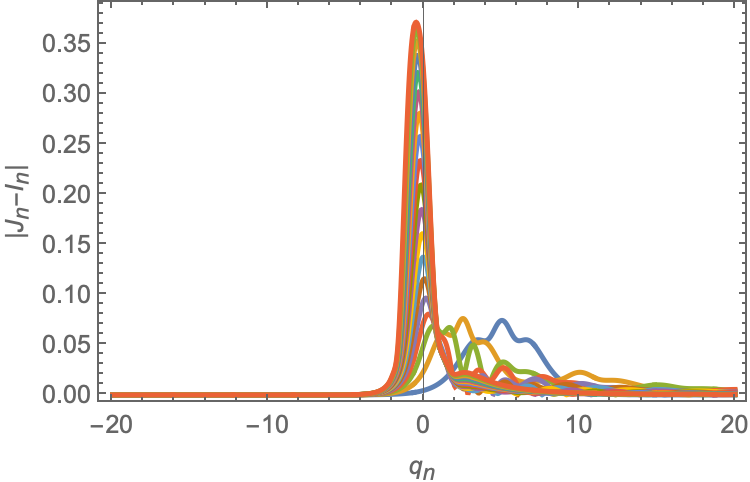}
        \caption{The modulus of the difference $J_n-\bar{J}_n$.}
    \end{subfigure}
    \caption{The $J_n$ functions (left) and the difference $J_n-\bar{J}_n$ (right) for $x_0=-5$, $\hbar=0.25$, $m=1$, $N=20$ for $n=2,\dots,20$ (blue, yellow, green, red, purple, brown, \dots).}\label{fig:Js}
\end{figure}

\bigskip
For the stitching method of the discretized path integral $G_N$ we first evaluate the $J_2,\dots,J_N$ integrals. The analytic continuation of the Rosen-Morse potential has an infinite set of singularities on the imaginary axis located at $x =i (1/2+k)\pi$ with $k \in \mathbb{Z}$. The convex hulls of these singularities in the upper and lower complex half planes consist of the lines $[i\pi/2, i \infty)$ and $(-i \infty,-i \pi/2]$. To avoid crossing singularities while deforming the integration domain of the $J_n$ integral, we identify the maximally allowed deformation angle
\begin{align}
    \theta_{max} = \tan^{-1} \left(\frac{\pi}{2\, \text{max}(|x_0|, |q_s|)}\right)\,.
\end{align}
To avoid the contour passing near any singularities we select an angle $0 < \theta < \text{min}( \theta_{max}, \pi/4)$ and evaluate the $J_n$ integral. In particular, we find that the choice $\theta = \frac{1}{2} \theta_{max}$ yields a well-behaved integral that allows for a quick numerical evaluation (see the left panel of \ref{fig:Js}). Alternatively, for large discretization $N$, we can use the eikonal approximation of the $J_n$ integral. For large $|q_n|$, the function $J_n$ is well approximated by the asymptotic $\bar{J}_n$ (see the right panel of \ref{fig:Js}). 

\begin{figure}
    \centering
    \begin{subfigure}[b]{0.49\textwidth}
        \includegraphics[width = \textwidth]{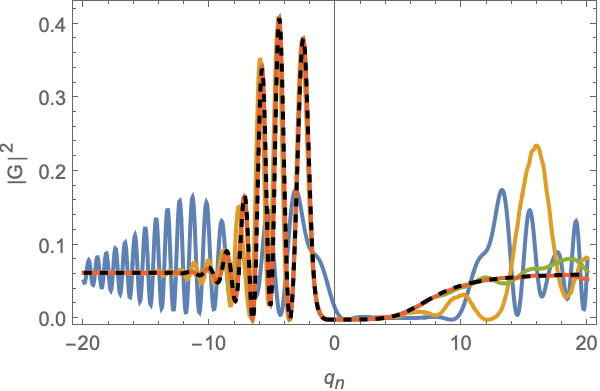}
        \caption{The modulus squared of the exact $|G(x_1,x_0,T)|^2$ and discretized propagator $|G_N(x_1,x_0,T)|^2$        .}
    \end{subfigure}
    \hfill
    \begin{subfigure}[b]{0.49\textwidth}
        \includegraphics[width = \textwidth]{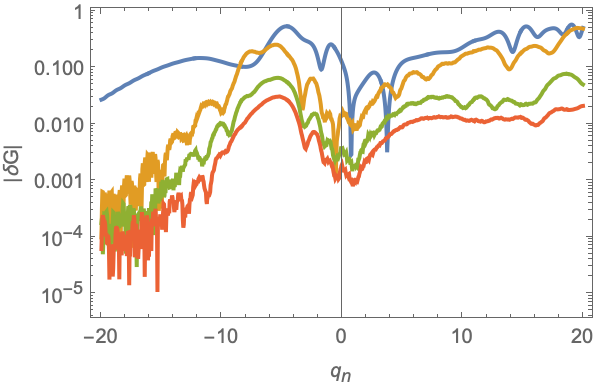}
        \caption{The modulus of the difference between the exact and the discretized propagator $|G_N(x_1,x_0,T) - G(x_1,x_0,T)|$.}
    \end{subfigure}
    \caption{Comparison of the exact propagator (black dashed) with the discretized path integral with $N=2,6,11,16$ (blue, yellow, green, red) for $x_0=-5$, $m=1$, $\hbar = 0.25$ and $V_0=1$.}\label{fig:slice}
\end{figure}

Next, we use a series of fast Fourier transformations to complete the stitching method. The resulting propagator for $x_0=-5, T=10,m=1, \hbar=1/4$ for the discretization levels $N=2,6, 11, 16$ (see the left panel of fig.\ \ref{fig:slice}) quickly approaches the continuum path integral as we refine the discretization (see the right panel of fig.\ \ref{fig:slice} for the modulus of the residual). In the interval $(-7,-2)$, the propagator consists of the interference of three classical paths interpolating between the initial $x_0$ and final state $x_1$ in the propagation time $T$. The interference of these three solutions leads to the oscillations in $|G|^2$. Outside this interval, a single real classical path contributes to the propagator leading to a smooth probability $|G|^2$.

\begin{figure}
    \centering
    \includegraphics[width = 0.5 \textwidth]{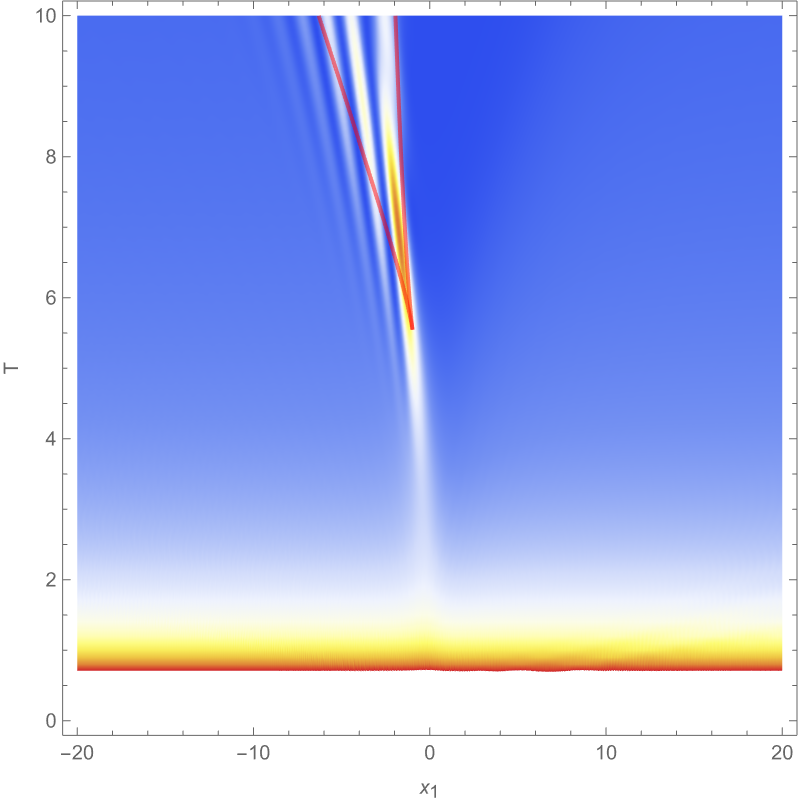}
    \caption{The time evolution of the propagator $|G(x_1,x_0,T)|^2$ starting at $x_0=-5$ as a function of the final position $x_1$ and propagation time $T$ for mass $m=1$ and reduced Planck constant $\hbar = 0.25$. This calculation was performed with the discretization level $N=100$. The caustic curves of the system are plotted in red.}\label{fig:evolution}
\end{figure}

The stitching method not only provides the final propagator as a function of the final position $x_1$ at time $T$ for fixed initial position $x_0$, but gives an estimate of the evolution of the propagator. In fig.\ \ref{fig:evolution}, we can observe the development of the propagator in time. First notice that the propagator decays as time increases. Next note that when the particle has enough time to interact with the barrier, the interference pattern consists of a series of stripes. For a more detailed analysis of the Rosen-Morse propagator as a function of time see \cite{Feldbrugge:2023_Rosen_Morse}.

\begin{figure}
    \centering
    \begin{subfigure}[b]{0.49\textwidth}
        \includegraphics[width = \textwidth]{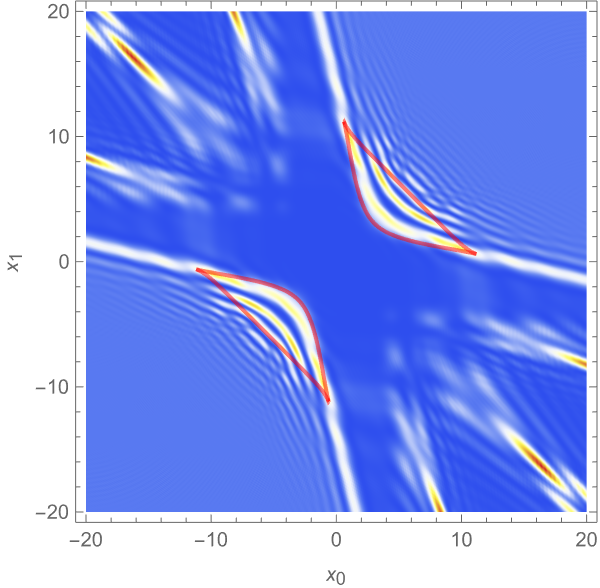}
        \caption{$N=5$.}
    \end{subfigure}
    \hfill
    \begin{subfigure}[b]{0.49\textwidth}
        \includegraphics[width = \textwidth]{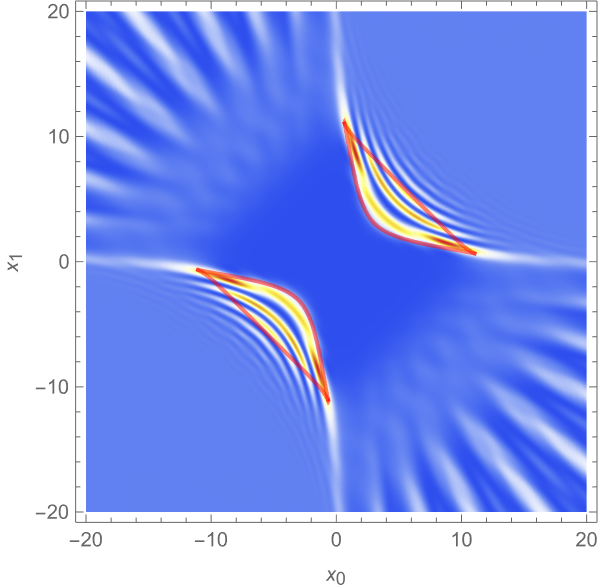}
        \caption{$N=10$.}
    \end{subfigure}\\
    \begin{subfigure}[b]{0.49\textwidth}
        \includegraphics[width = \textwidth]{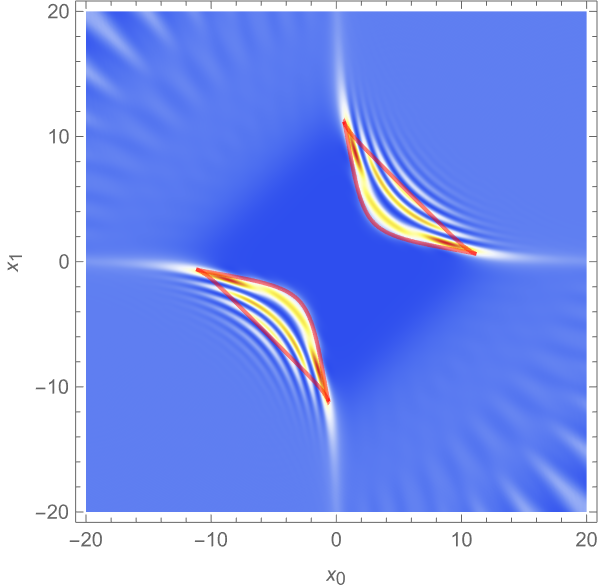}
        \caption{$N=15$.}
    \end{subfigure}
    \hfill
    \begin{subfigure}[b]{0.49\textwidth}
        \includegraphics[width = \textwidth]{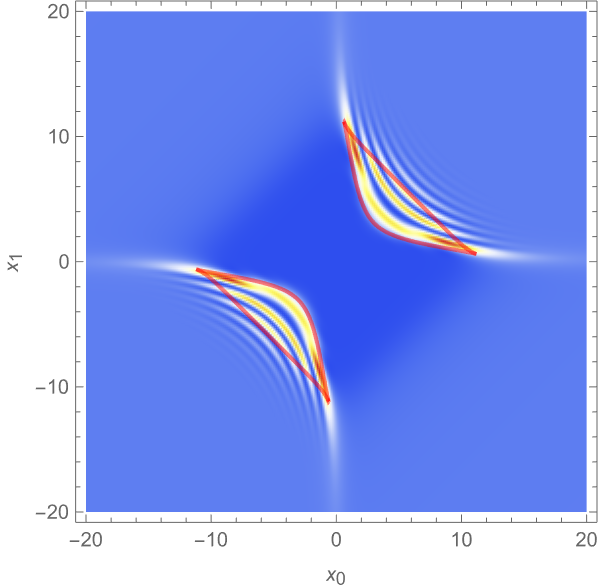}
        \caption{Exact propagator}
    \end{subfigure}
    % \begin{subfigure}[b]{0.49\textwidth}
    %     \includegraphics[width = \textwidth]{G_field_N=20.png}
    %     \caption{$N=20$.}
    % \end{subfigure}
    \caption{Comparison of the exact propagator with the discretized path integral with $N=5,10,15$ for $m=1$, $\hbar = 0.25$ and $V_0=1$. The caustic curves of the continuum theory are overlaid (red).}\label{fig:field}
\end{figure}

\begin{figure}
    \centering
    \begin{subfigure}[b]{0.49\textwidth}
        \includegraphics[width = \textwidth]{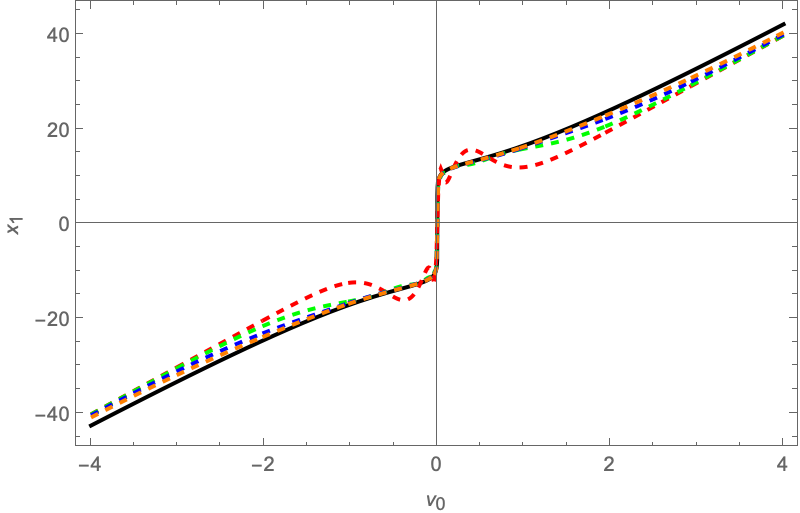}
        \caption{$x_0=0$}
    \end{subfigure}
    \hfill
    \begin{subfigure}[b]{0.49\textwidth}
        \includegraphics[width = \textwidth]{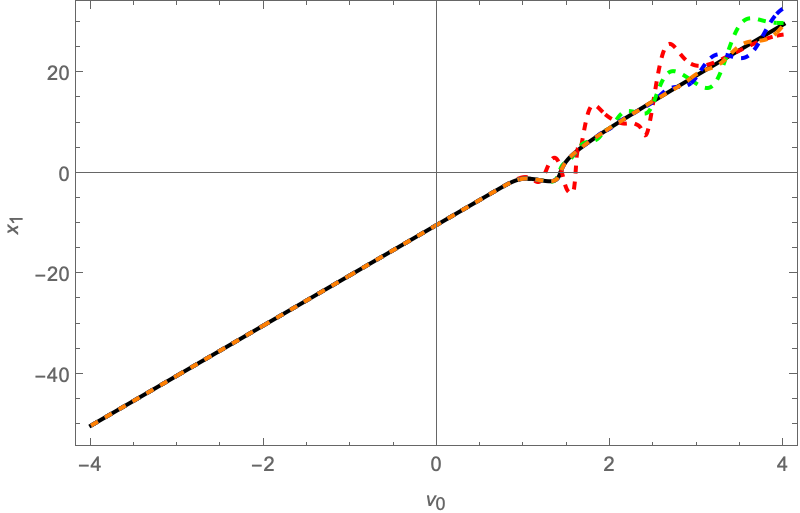}
        \caption{$x_0=-10$}
    \end{subfigure}
    \caption{Comparison of the continuum classical paths (black) with the discretized classical paths  $N=5,10,15$ (red, green, and blue) for $m=1$ and $V_0=1$ with $q_1= x_0 + v_0 a$.}\label{fig:saddle}
\end{figure}

\bigskip
When evaluating the propagator for a lattice of initial positions $x_0$, we recover the complete interference pattern of the discretized path integral $G_N$ (see fig.\ \ref{fig:field}). For small discretization, the interference pattern nicely matches the exact propagator for small $x_0$ and $x_1$. Artefacts are visible in the upper left and lower right quadrants in fig. \ref{fig:field}. Consider for example the stripes in the upper left panel of fig.\ \ref{fig:field}. These artefacts are not caused by the stitching method but are rather a mismatch between the discretized path integral $G_N$ and the continuum path integral $G$ at finite $N$. To see this, take the gradient of the exponent of the discretized path integral with respect to the integration variables, $\nabla_{\bm{q}_n} \sum_{n=0}^{N-1} S_n$ for $n=1,\dots,N-1$. The resulting tower of algebraic equations is solved iteratively with the rule
\begin{align}
    q_n = -q_{n-2} + 2 q_{n-1} - \frac{a^2}{m} \nabla V(q_{n-1}) \,,
\end{align}
for $n=2,\dots, N$. Starting with the initial position $q_0 = x_0$ we can map the position of the particle $q_1$ at time $t=a$ to the final position $x_1=q_N$, yielding the map $\xi(q_1)$ for fixed $x_0$ (for the map $\xi$ for $x_0=0$ and $x_0=-10$ see fig.\ \ref{fig:saddle}). The propagator spikes at the caustics of the discretized path integral $G_N$, when the inverse $\xi(q_1)$ changes in number of real solutions. Hence, we solve the equation $\xi(q_1)=x_1$ for $q_1$ and look for $x_1$ for which the number of real solutions changes. For $x_0=0$, the discretized classical path is relatively close to the continuum classical path. For $x_0=-10$ we observe several oscillations in $\xi$ for positive $q_1$. After the particle has crossed the barrier, the final position as a function of the initial velocity (or $q_1$) for the discretized path oscillates, unlike the continuum version. The number inverses in $\xi^{-1}(x_1) = \{q_1 | \xi(q_0) = x_1\}$ (geometrically corresponding to the intersection of the curve with horizontal lines) changes at the maxima and minima of $\xi(q_0)$. These are caustics and precisely correlate with the stripes in the upper left and lower right quadrants of the interference patterns observed in fig.\ \ref{fig:field}. As we refine the discretization, the artifacts move to larger $|x_0|$ and $|x_1|$, and the discretized propagator approaches the exact real-time path integral (see the remaining panels in fig.\ \ref{fig:field}).

\bigskip
The interference pattern in the continuum path integral is bounded by caustics (the red curves in fig.\ \ref{fig:field}). In the continuum theory, saddle points of the path integral are solutions to the equation of motion
\begin{align}
    m \ddot{x} = -\nabla V(x)
\end{align}
with the boundary conditions $x(t=0) = x_0$ and $x(t=T) = x_1$. The caustics are points in the $x_0$-$x_1$ plane for which the number of real solutions to this boundary value problem changes. To evaluate the caustic curves, we follow two steps. First, we evaluate the critical curves using the initial value problem $x(0)=x_0$, $\dot{x}(0) = v_0$, 
\begin{align}
    \mathcal{C} = \{(x_0,v_0)\, |\, \det \nabla_{v_0} x(T)  = 0\}\,.
\end{align}
We use a vector notation to suggest the extension to higher-dimensional quantum mechanics. Next, we evaluate the caustic curve by mapping the critical curve in the initial position velocity plane to the initial and final position plane,
\begin{align}
    x(\mathcal{C}) = \{(x_0,x_1)\, |\, (x_0,v_0) \in \mathcal{C} \text{ and }x(T) = x_1\}\,.
\end{align}
Inside the caustic regions, there exist three real solutions to the boundary value problem (one direct and two bouncing solutions), reflected in the oscillations in the modulus squared propagator $|G(x_1,x_0,T)|^2$. Outside the caustic regions, there exists only a single real classical path interpolating between $x_0$ and $x_1$ resulting in a propagator with fewer oscillations. The oscillations outside the caustic regions can be interpreted as the contribution of a relevant complex classical path. For more details on the role of caustics in quantum mechanics, we refer to \cite{Schulman:1975, Schulman:2012, Feldbrugge:2023_Rosen_Morse}.

%%%%%%%%%%%%%%%%%%%%%%%%%%%%%%%
\subsection{A smooth step}\label{sec:tanh}
In the previous section, we considered the evolution of a particle encountering a barrier. The potential vanishes for $x\to \pm \infty$. Next, consider the path integral for a particle encountering a step potential, where the potential at $x \to -\infty$ and $x \to \infty$ differ by a nonzero value $V_0$. In particular, we consider the smooth step potential
\begin{align}
    V(x) = \frac{V_0}{2}\left[1 + \tanh x\right]\,,
\end{align}
with positive strength $V_0$, smoothly interpolating between $0$ at $x \to -\infty$ to $V_0$ at $x\to \infty$. 

\begin{figure}
    \centering
    \begin{subfigure}[b]{0.32\textwidth}
        \includegraphics[width = \textwidth]{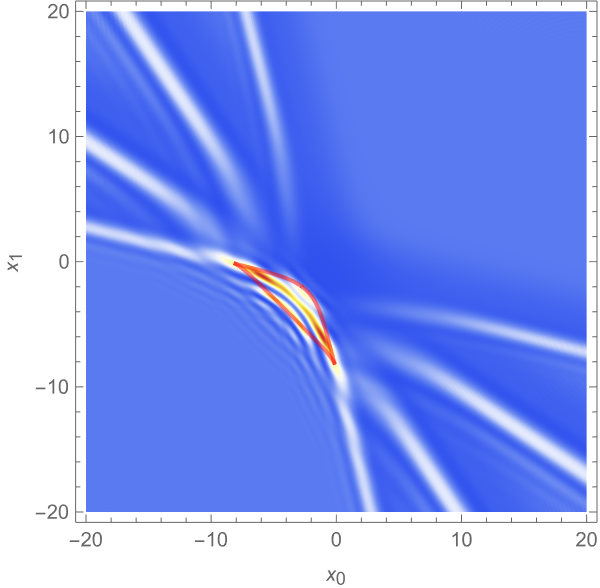}
        \caption{$N=5$.}
    \end{subfigure}
    \hfill
    \begin{subfigure}[b]{0.32\textwidth}
        \includegraphics[width = \textwidth]{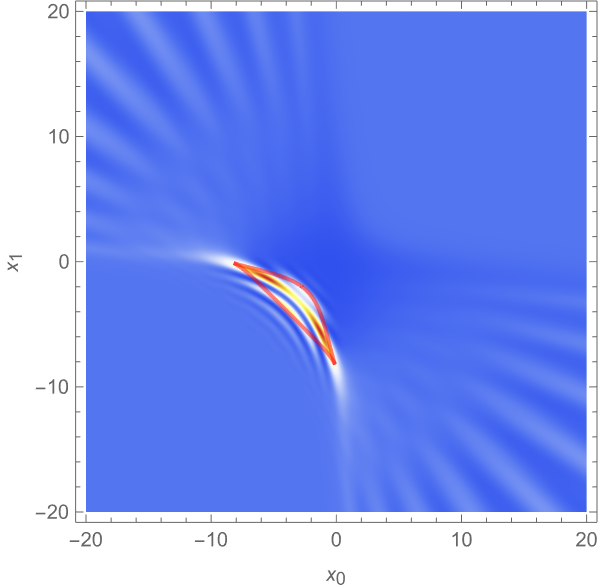}
        \caption{$N=10$.}
    \end{subfigure}
    \hfill
    % \begin{subfigure}[b]{0.32\textwidth}
    %     \includegraphics[width = \textwidth]{G_tanh_field_N=15.png}
    %     \caption{$N=15$.}
    % \end{subfigure}
    % \hfill
    \begin{subfigure}[b]{0.32\textwidth}
        \includegraphics[width = \textwidth]{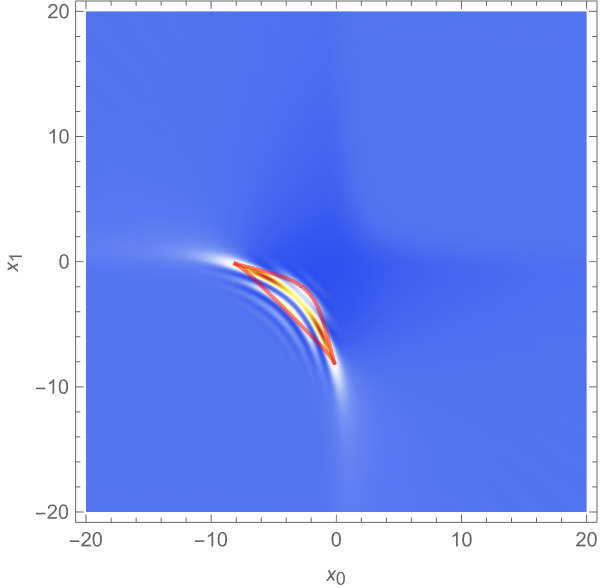}
        \caption{$N=20$.}
    \end{subfigure}
    \caption{The discretized path integral for the smooth step potential with discretization $N=5,10, 20$ for $m=1$, $\hbar = 0.25$ and $V_0=1$, with the caustic curves of the continuum theory (red).}\label{fig:field_tanh}
\end{figure}

As the analytic continuation of the $\tanh$ function has singularities at $x =i (1/2+k)\pi$ with $k \in \mathbb{Z}$, not unlike the Rosen-Morse potential, we use the same deformation of the $J_n$ integral. The resulting propagator $G_N$ for $x_0,x_1 <0$ qualitatively resembles the propagator of the Rosen-Morse potential (see fig.\ \ref{fig:field_tanh}). For positive $x_0$ and $x_1$ the propagator $|G_N(x_0,x_1,T)|^2$ is relatively smooth as there only exists a single real classical path in the continuum theory. Note that there are a series of artefacts in the upper left and lower right quadrants of fig.\ \ref{fig:field_tanh} for small discretization $N$. As $N$ is increased these artifacts are pushed away from the central region. 

%%%%%%%%%%%%%%%%%%%%%%%%%%%%%%%
\subsection{The double-well potential}\label{sec:double-well}
Finally, we consider the path integral for a non-relativistic particle in a double-well
\begin{align}
    V(x) = (x^2-1)^2\,.
\end{align}
As the potential diverges to $+\infty$ as $|x| \to \infty$, the path integral no longer evaluates to a function but rather a distribution. To see this, first note that there exist an infinite number of solutions to the classical boundary value problem 
\begin{align}
    m \ddot{x} = - V'(x)\,, \quad x(0)=x_0\,, \quad x(T)=x_1\,,
\end{align}
labeled by the number of oscillations the particle performs before reaching its destination. Each classical path is included in the path integral. Their quantum representation can be recovered through the convolution of an initial state
\begin{align}
    \psi_t(x_1) = \int \psi_0(x_0) G(x_1,x_0,t)\mathrm{d}x_0\,,
\end{align}
for $0\leq t \leq T$, where the initial position $x_0=x(0)$ is filtered by the modulus of the initial state $|\psi_0|^2$ and the initial momentum $p_0= m \dot{x}(0)$ is selected through the variation of the phase. For a Gaussian initial state, 
\begin{align}
    \psi_0(x_0) = \frac{1}{\sqrt[4]{2 \pi \sigma^2}}e^{-\frac{(x_0-\mu)^2}{4 \sigma^2} + i p_0 \cdot x_0/ \hbar}
\end{align}
we see this through the mean position $\mu$, the spread $\sigma$ and the mean initial momentum $p_0$. Now, as the initial momenta of the classical paths interpolating between $x_0$ and $x_1$ in propagation time $T$ are unbounded from above, the Fourier transform of the propagator with respect to $x_0$
\begin{align}
    F(k) = \int G(x_1,x_0,T) e^{i k x_0}\mathrm{d}x_0
\end{align} 
does not decay for large frequencies but rather spikes at the initial momenta of these classical paths. The propagator is ``infinitely oscillatory'' in $x_0$. More rigorously, as the Fourier transform is not dominated by $K/|k|$ for large frequencies $|k|$ for some positive constant $K$, the propagator $G(x_1,x_0,T)$ is discontinuous as a function of $x_0$. By symmetry, the same applies for the final position $x_1$. This property is nicely explained for a particle on a box in \cite{Fulling:2003}. For a more general discussion on this observation see \cite{Feldbrugge:2023_Existence}. For confined theories, the discretized path integral $G_N(x_1,x_0,T)$ does not converge to a number in the limit $N\to \infty$. In appendix \ref{ap:gauss}, we discuss the application of the stitching method to the evolution of a Gaussian initial state, which can be used in the case of a double-well potential. However, there exist more efficient methods to evolve the time dependent Schrödinger equation, including the Crank-Nicolson method \cite{Crank:1947} and the Suzuki-Trotter method \cite{De_Raedt:1987}. Note that the propagator of the Rosen-Morse potential does not suffer from this problem, as there are always a finite number of solutions to any boundary value problem.

\begin{figure}
    \centering
    \includegraphics[width=0.5\textwidth]{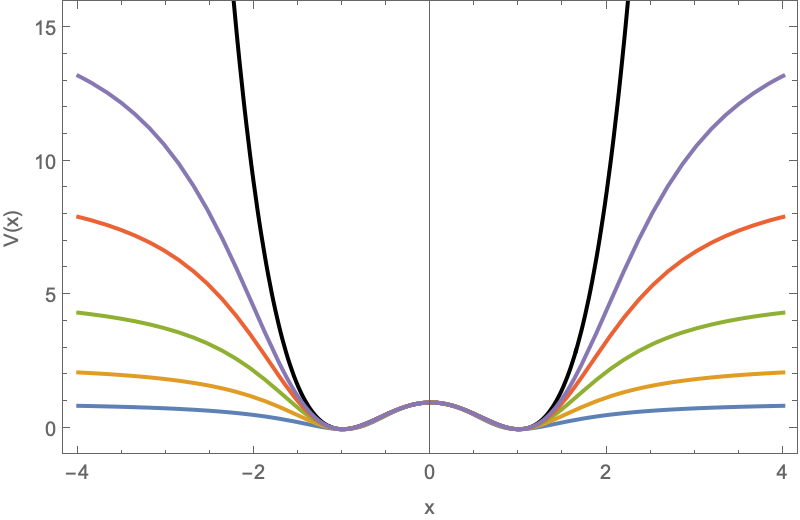}
    \caption{The truncated double-well potential $V(x) = (x^2-1)^2/(1+(x/\alpha)^4)$ for $\alpha =1, 1.25, 1.5, 1.75, 2$ (blue, yellow, green, red and purple) compared with the double-well potential $V(x)=(x^2-1)^2$ (black)}\label{fig:double-well_reg}
\end{figure}

\begin{figure}
    \centering
    \begin{subfigure}[b]{0.32\textwidth}
        \includegraphics[width = \textwidth]{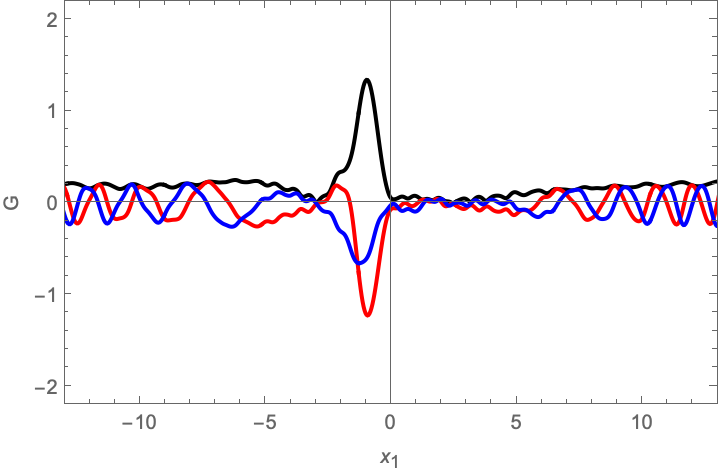}
        \caption{$\alpha=1$.}
    \end{subfigure}
    \hfill
    \begin{subfigure}[b]{0.32\textwidth}
        \includegraphics[width = \textwidth]{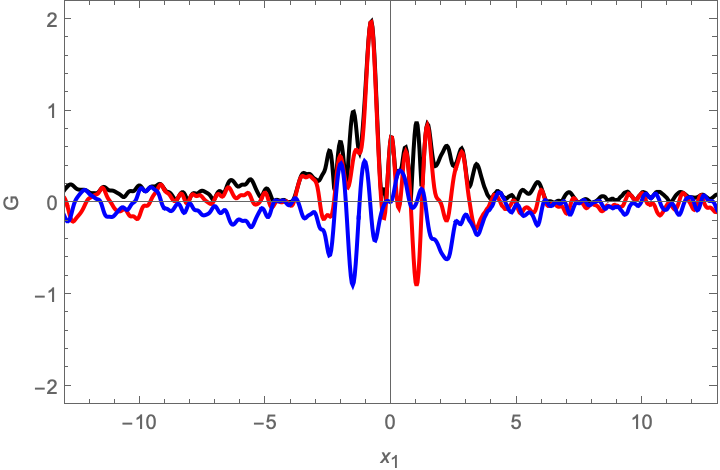}
        \caption{$\alpha=1.5$.}
    \end{subfigure}
    \hfill
    \begin{subfigure}[b]{0.32\textwidth}
        \includegraphics[width = \textwidth]{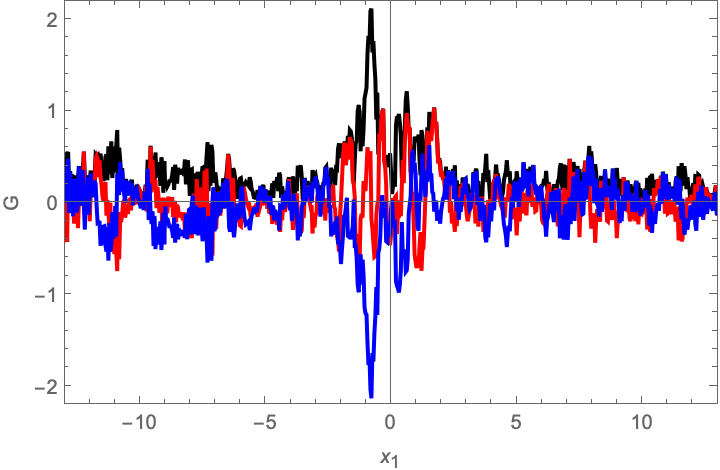}
        \caption{$\alpha=2$.}
    \end{subfigure}
    \caption{The truncated double-well path integral (real part in red, imaginary part in blue, and the magnitude in black) for $\alpha=1,1.5,$ and $2$ as a function of the final position, with initial position $x_0=-1$, mass $m=1$, Planck constant $\hbar=0.25$, and discretization level $N=20$.}\label{fig:double-well_alpha}
\end{figure}

The stitching method discussed in section \ref{sec:stitching} does not directly apply to the path integral of the double-well potential as the potential diverges faster than a quadratic for large $|x|$. To circumvent this problem, we truncate the double-well potential as 
\begin{align}
    V(x) = \frac{(x^2-1)^2}{1 + (x/\alpha)^4}\,,
\end{align}
which approaches $\alpha^4$ for $|x| \to \infty$. The traditional double-well potential is recovered in the limit $\alpha \to \infty$ (see fig.\ \ref{fig:double-well_reg}). 
For finite $\alpha$ the stitching method converges to a smooth but wildly oscillatory function. As $\alpha$ is increased the propagator becomes increasingly oscillatory (see fig.\ \ref{fig:double-well_alpha}). We can trace this behaviour back to the $J_n$ functions. As $N$ increases, the effective Planck constant \eqref{eq:eff_Planck} decreases while the effective phase variation \eqref{eq:eff_phase} remains significant as the potential assumes increasingly large values. For large $\alpha$, the $J_n$ integral for large $n \leq N$ becomes increasingly oscillatory, signalling the nature of the propagator $G(x_1,x_0,T)$.

\begin{figure}
    \centering
    \begin{subfigure}[b]{0.32\textwidth}
        \includegraphics[width = \textwidth]{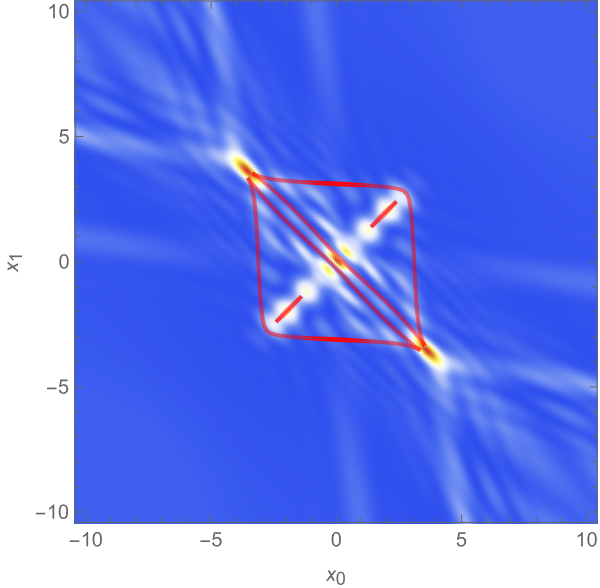}
        \caption{$N=5$.}
    \end{subfigure}
    \hfill
    \begin{subfigure}[b]{0.32\textwidth}
        \includegraphics[width = \textwidth]{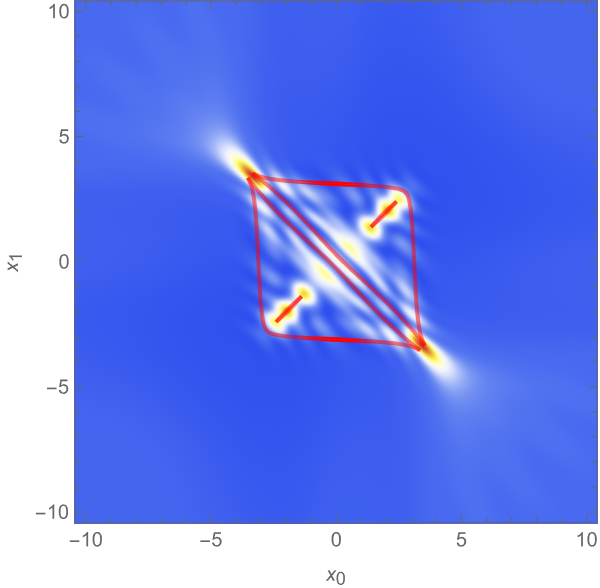}
        \caption{$N=10$.}
    \end{subfigure}
    \hfill
    % \begin{subfigure}[b]{0.32\textwidth}
    %     \includegraphics[width = \textwidth]{G_double-well_field_N=15.png}
    %     \caption{$N=15$.}
    % \end{subfigure}
    % \hfill
    \begin{subfigure}[b]{0.32\textwidth}
        \includegraphics[width = \textwidth]{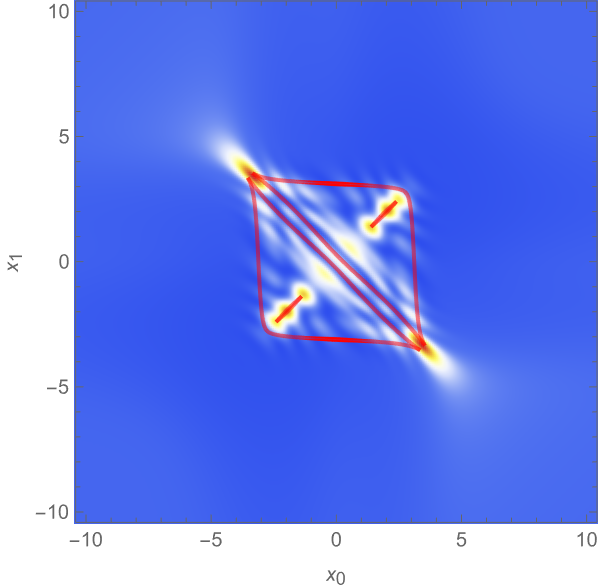}
        \caption{$N=20$.}
    \end{subfigure}
    \caption{The modulus squared value of the propagator of the Gaussian truncated double-well potential, with mass $m=1$, reduced Planck constant $\hbar =0.25$ and propagation time $T=10$. The caustics of the continuum double-well potential are shown in red.}\label{fig:field_double-well}
\end{figure}

Another interesting way to regulate the double-well potential is through the use of Gaussian functions. In particular, we can model the double-well potential in terms of two Gaussian functions
\begin{align}
    V(x) = - \left[e^{-\left(\frac{x - 2}{2}\right)^2} + e^{-\left(\frac{x + 2}{2}\right)^2}\right]\,,
\end{align}
This truncated potential includes the physics of quantum tunneling while its path integral remains a function. For the evaluation of the $J_n$ integrals, note that the analytic continuation of the Gaussian function does not have singularities at finite distance in the complex plane. In this paper, we evaluate the $J_n$ integral with a constant deformation parametrized by selecting a constant angle $\theta < \pi/4$. The resulting interference pattern nicely matches the caustics of the continuum theory (see fig.\ \ref{fig:field_double-well}). Our method applies to any potential written as the sums of Gaussian functions, enabling the numerical study of the path integral for a large class of physical potentials.

%%%%%%%%%%%%%%%%%%%%%%%%%%%%%%%
\section{Relativistic quantum systems}\label{sec:relativistic}
For relativistic theories, the world-line quantization is captured by the path integral 
\begin{align}
    G(x_1,x_0) = \int_0^\infty \left[\int_{x(0)=x_0}^{x(\tau)=x_1} e^{i S[x,\tau]}\mathcal{D}x\right] \mathrm{d}\tau\,,
\end{align}
interpreted as the amplitude to propagate from $x_0$ to $x_1$, with the lapse or Schwinger time $\tau$ and the action $S$. The action of a relativistic particle in Minkowski space interacting with a potential $V$ assumes the form
\begin{align}
    S[x,\tau] = \int_0^1 \left[\frac{m \dot{x}^2}{2\tau} - \tau V(x)\right]\mathrm{d}t\,.
\end{align}
For a relativistic particle in spacetime, the boundary conditions $x_0$ and $x_1$ are spacetime events and the path $x(t)$ is a word-line in spacetime interpolating between these two spacetime events. In quantum gravity, following quantum geometrodynamics \cite{Wheeler:1957}, $x_0$ and $x_1$ represent two three-geometries and the path $x(t)$ represents a spacetime manifold interpolating between these three-geometries. In quantum cosmology, the variable $x$ is often identified with the scale factor of our universe.

The stitching method for the evolution of path integrals directly applies to the inner integral 
\begin{align}
    G(x_1,x_0,\tau) = \int_{x(0)=x_0}^{x(\tau)=x_1} e^{i S[x,\tau]}\mathcal{D}x\,.
\end{align}
We can implement the $\tau$ integral by deforming the integration domain $(0,\infty)$ for $\tau$ into the complex plane to improve the convergence of the $\tau$ integral. For small $|\tau|$, the action is dominated by the kinetic term and a deformation along the imaginary axis is optimal. For large $|\tau|$, the action is dominated by the potential $V$. By studying the structure of the analytic continuation of $V$ at complex infinity, we can identify a reasonable deformation of the integration domain. This procedure is more efficient than the scheme discussed above, and will be discussed in detail in an upcoming paper. For an example of an optimal deformation of the $\tau$ integral in quantum cosmology see \cite{Feldbrugge:2017}.

It may be possible to extend the stitching methods to the path integral over quantum fields, 
\begin{align}
    \int e^{i S[\phi]/\hbar} \mathcal{D} \phi\,.
\end{align}
However, the extension is not directly obvious as the subsequential coupling of integration variables in the discretized path integral in quantum mechanics is replaced by a network of coupled integration variables. This extension will be studied in future explorations of real-time quantum evolution.

%%%%%%%%%%%%%%%%%%%%%%%%%%%%%%%
\section{Implementation and performance}\label{sec:performance}
We now briefly discuss the implementation of the stitching method and its performance.
\subsection{Implementation}
\noindent We implement the stitching method in four steps:
\begin{enumerate}
    \item We define a regular lattice spatial lattice and approximate/evaluate the $J_n$ integrals on this lattice.
    \item We evaluate the residual of $J_n$ with respect to $\bar{J}_n$, defined as $\delta J_n = J_n - \bar{J}_n$. This residual will generally vanish for large $\lVert\bm{q}_n\rVert$.
    \item As the convolutions will often grow beyond the size of the original lattice, we extend the lattice and pad the residual $\delta J_n$ with zeros. The required padding depends on the physical problem in question. 
    \item Next, we evaluate the convolutions using a series of fast Fourier transformations to obtain the discretized path integral $G_N$. When required, we downsize the result to match the original lattice.
\end{enumerate}
In the examples presented in this paper, we evaluated the $J_n$ integrals both numerically and approximated them using the eikonal approximation. The two methods performed comparably for one-dimensional path integrals. We expect the eikonal approximation to be the more efficient option in high-dimensional quantum systems. 

The stitching method was implemented in Julia and the $J_n$ integrals were performed in parallel. 

\subsection{Convergence}
\begin{figure}
    \centering
    \includegraphics[width = 0.6\textwidth]{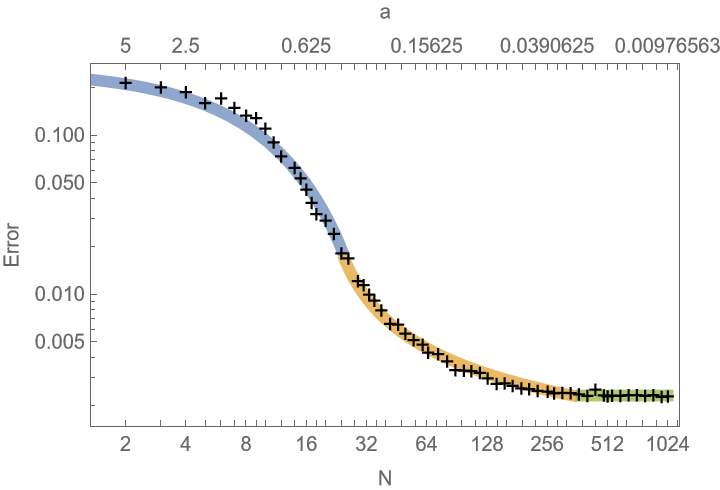}
    \caption{The convergence of the discretized propagator to the exact result as a function of $N$, ranging from 2 to 1024. The corresponding lattice spacing $a$ has also been included along the top. The parameters used in this plot were $T=10$, $\hbar=0.25$, $x_0=-5$, and $m=1$, with the bounds $A=-50$ and $B=50$. We highlight an exponential convergence (blue), a logarithmic convergence (orange), and a region where the error levels off (green).}\label{fig:rate}
\end{figure}

To investigate the convergence of the discretized path integral to the continuum result, we evaluate the propagator of the Rosen-Morse system for a set of 61 discretization sizes roughly logarithmically equispaced between $N=2$ and $N=1024$, and compare the resulting propagator to the exact path integral with the error 
\begin{align}
    \epsilon_N = \frac{1}{B-A}\int_{A}^{B} \left|G(x_1,x_0,T) - G_N(x_1,x_0,T)\right|\mathrm{d}x_1\,.
\end{align}
We can identify three major regimes in the convergence of the discretized path integral (see fig.\ \ref{fig:rate}). Initially, the convergence is exponential. There is then a period of logarithmic convergence, until the error seems to level off. Whether it genuinely levels off at a constant value, or is simply decreasing slowly, requires more data to discern. The exponential and logarithmic curve scale as roughly $e^{-0.1 N}$ and $\ln (0.05 N)^{-1}$ respectively. One should certainly not read into these scalings, they will be parameter dependent, and we include them only for completeness.

For $N=10$ ($a=0.5$), we generally obtained percent level accuracy for the Rosen-Morse barrier system. Problems with steeper potentials will require higher discretization levels to obtain accurate results.

\subsection{Performance}
The approximation/evaluation of the $J_n$ integral is typically the bottleneck of the stitching method. Luckily the $J_n$ integrals are evaluated independently and are trivially parallelizable. The evaluation of the propagator for a fixed initial position $x_0$ and a reasonable number of discretization times $N=20$ evaluates in a couple of seconds on a modern laptop with a 2.4 GHz 8-Core Intel Core i9 processor. The propagators as a function of initial position $x_0$, final position $x_1$, and propagation time $T$ on a regular lattice evaluates in a couple of minutes. None of the calculations presented in this paper require high-performance computing facilities such as a computer cluster.

%%%%%%%%%%%%%%%%%%%%%%%%%%%%%%%
\section{Conclusion}\label{sec:conclusion}
For theories where the potential is dominated by a quadratic at infinity, we rewrite the time discretized $d(N-1)$-dimensional integrals path integral in terms of $N-1$ $d$-dimensional oscillatory integrals that are subsequently stitched together using $2(N-1)$ $d$-dimensional fast Fourier transformations. The method provides an efficient way to numerically evaluate the quantum mechanical propagator $G(\bm{x}_1,\bm{x}_0,T)$ for fixed initial position $\bm{x}_0$ for a lattice of final positions $\bm{x}_1$ and propagation times $T$. We demonstrate the method for a Rosen-Morse barrier, a smooth step and a double-well potential, and compare the resulting interference pattern with the caustics of the corresponding classical theory. We discuss the application of this method to the word-line quantization in relativistic quantum physics and hope that this method will find applications in many fields including quantum field theory and quantum gravity. 

As the Feynman propagator is the building block of many concepts in quantum physics, the method is directly applicable to, for example, the evaluation of expectation values 
\begin{align}
    \langle Q(\bm{x}(t^*)) \rangle 
    &= \frac{\int_{\bm{x}(0)=\bm{x}_0}^{\bm{x}(T) = \bm{x}_1} Q(\bm{x}(t^*)) e^{i S[\bm{x}]/\hbar} \mathcal{D}\bm{x}}{\int_{\bm{x}(0) = \bm{x}_0}^{\bm{x}(T) = \bm{x}_1} e^{i S[\bm{x}]/\hbar} \mathcal{D}\bm{x}}\\
    &= \frac{\int G(\bm{x}_1,\bm{x}^*,T-t^*) Q(\bm{x}^*) G(\bm{x}^*,\bm{x}_0,t^*)\mathrm{d}\bm{x}^*}{G(\bm{x}_1,\bm{x}_0,T)}\,,
\end{align}
of the observable $Q(\bm{x}(t^*))$ measured at the time $0<t^*<T$, and correlation functions such as the unequal time two-point function $\langle Q_1(\bm{x}(t_1))Q_2(\bm{x}(t_2))\rangle$.

The stitching method enables the study of a large variety of problems in real-time quantum physics that are otherwise very expensive to approach. In particular, in the near future, we plan to apply the stitching method to multi-particle path integrals focusing on scattering experiments, the path integral quantization of fermions, and quantum cosmological models of the early universe.

In an upcoming paper, we plan to extend these methods to more general potentials in quantum mechanics and the path integral over quantum fields. 

\section*{Acknowledgments}
The work of JF is supported by the STFC Consolidated Grant ‘Particle Physics at the Higgs Centre,’ and, respectively, by a Higgs Fellowship and the Higgs Chair of Theoretical Physics at the University of Edinburgh. JJ is supported by Science Foundation Ireland under Grant number 22/PATH-S/10704.

For the purpose of open access, the authors have applied a Creative Commons Attribution (CC BY) license to any Author Accepted Manuscript version arising from this submission.

\bibliographystyle{apsrev4-1}
\bibliography{library.bib}

\begin{thebibliography}{46}%
\makeatletter
\providecommand \@ifxundefined [1]{%
 \@ifx{#1\undefined}
}%
\providecommand \@ifnum [1]{%
 \ifnum #1\expandafter \@firstoftwo
 \else \expandafter \@secondoftwo
 \fi
}%
\providecommand \@ifx [1]{%
 \ifx #1\expandafter \@firstoftwo
 \else \expandafter \@secondoftwo
 \fi
}%
\providecommand \natexlab [1]{#1}%
\providecommand \enquote  [1]{``#1''}%
\providecommand \bibnamefont  [1]{#1}%
\providecommand \bibfnamefont [1]{#1}%
\providecommand \citenamefont [1]{#1}%
\providecommand \href@noop [0]{\@secondoftwo}%
\providecommand \href [0]{\begingroup \@sanitize@url \@href}%
\providecommand \@href[1]{\@@startlink{#1}\@@href}%
\providecommand \@@href[1]{\endgroup#1\@@endlink}%
\providecommand \@sanitize@url [0]{\catcode `\\12\catcode `\$12\catcode `\&12\catcode `\#12\catcode `\^12\catcode `\_12\catcode `\%12\relax}%
\providecommand \@@startlink[1]{}%
\providecommand \@@endlink[0]{}%
\providecommand \url  [0]{\begingroup\@sanitize@url \@url }%
\providecommand \@url [1]{\endgroup\@href {#1}{\urlprefix }}%
\providecommand \urlprefix  [0]{URL }%
\providecommand \Eprint [0]{\href }%
\providecommand \doibase [0]{http://dx.doi.org/}%
\providecommand \selectlanguage [0]{\@gobble}%
\providecommand \bibinfo  [0]{\@secondoftwo}%
\providecommand \bibfield  [0]{\@secondoftwo}%
\providecommand \translation [1]{[#1]}%
\providecommand \BibitemOpen [0]{}%
\providecommand \bibitemStop [0]{}%
\providecommand \bibitemNoStop [0]{.\EOS\space}%
\providecommand \EOS [0]{\spacefactor3000\relax}%
\providecommand \BibitemShut  [1]{\csname bibitem#1\endcsname}%
\let\auto@bib@innerbib\@empty
%</preamble>
\bibitem [{\citenamefont {{Feynman}}(2005)}]{Feynman:2005}%
  \BibitemOpen
  \bibfield  {author} {\bibinfo {author} {\bibfnamefont {R.~P.}\ \bibnamefont {{Feynman}}},\ }\href {\doibase 10.1142/5852} {\emph {\bibinfo {title} {{Feynman's Thesis - a New Approach to Quantum Theory}}}},\ edited by\ \bibinfo {editor} {\bibfnamefont {L.~M.}\ \bibnamefont {{Brown}}}\ (\bibinfo {year} {2005})\BibitemShut {NoStop}%
\bibitem [{\citenamefont {{Feynman}}(1948)}]{Feynman:1948}%
  \BibitemOpen
  \bibfield  {author} {\bibinfo {author} {\bibfnamefont {R.~P.}\ \bibnamefont {{Feynman}}},\ }\href {\doibase 10.1103/RevModPhys.20.367} {\bibfield  {journal} {\bibinfo  {journal} {Reviews of Modern Physics}\ }\textbf {\bibinfo {volume} {20}},\ \bibinfo {pages} {367} (\bibinfo {year} {1948})}\BibitemShut {NoStop}%
\bibitem [{\citenamefont {{Feynman}}(1985)}]{Feynman:1985}%
  \BibitemOpen
  \bibfield  {author} {\bibinfo {author} {\bibfnamefont {R.~P.}\ \bibnamefont {{Feynman}}},\ }\href@noop {} {\emph {\bibinfo {title} {{QED. The strange theory of light and matter}}}}\ (\bibinfo {year} {1985})\BibitemShut {NoStop}%
\bibitem [{\citenamefont {Feynman}\ and\ \citenamefont {Hibbs}(1965)}]{Feynman:1965}%
  \BibitemOpen
  \bibfield  {author} {\bibinfo {author} {\bibfnamefont {R.~P.}\ \bibnamefont {Feynman}}\ and\ \bibinfo {author} {\bibfnamefont {A.~R.}\ \bibnamefont {Hibbs}},\ }\href@noop {} {\emph {\bibinfo {title} {{Quantum mechanics and path integrals}}}},\ International series in pure and applied physics\ (\bibinfo  {publisher} {McGraw-Hill},\ \bibinfo {address} {New York, NY},\ \bibinfo {year} {1965})\BibitemShut {NoStop}%
\bibitem [{\citenamefont {Schulman}(2012)}]{Schulman:2012}%
  \BibitemOpen
  \bibfield  {author} {\bibinfo {author} {\bibfnamefont {L.}~\bibnamefont {Schulman}},\ }\href@noop {} {\emph {\bibinfo {title} {Techniques and Applications of Path Integration}}},\ Dover Books on Physics\ (\bibinfo  {publisher} {Dover Publications},\ \bibinfo {year} {2012})\BibitemShut {NoStop}%
\bibitem [{\citenamefont {Kleinert}(2004)}]{Kleinert:2004}%
  \BibitemOpen
  \bibfield  {author} {\bibinfo {author} {\bibfnamefont {H.}~\bibnamefont {Kleinert}},\ }\href {https://cds.cern.ch/record/788154} {\emph {\bibinfo {title} {{Path integrals in quantum mechanics, statistics, polymer physics, and financial markets; 3rd ed.}}}}\ (\bibinfo  {publisher} {World Scientific},\ \bibinfo {address} {River Edge, NJ},\ \bibinfo {year} {2004})\ \bibinfo {note} {based on a Course on Path Integrals, Freie Univ. Berlin, 1989/1990}\BibitemShut {NoStop}%
\bibitem [{\citenamefont {{Zee}}(2003)}]{Zee:2003}%
  \BibitemOpen
  \bibfield  {author} {\bibinfo {author} {\bibfnamefont {A.}~\bibnamefont {{Zee}}},\ }\href@noop {} {\emph {\bibinfo {title} {{Quantum field theory in a nutshell}}}}\ (\bibinfo {year} {2003})\BibitemShut {NoStop}%
\bibitem [{\citenamefont {{Polchinski}}(1998)}]{Polchinski:1998}%
  \BibitemOpen
  \bibfield  {author} {\bibinfo {author} {\bibfnamefont {J.}~\bibnamefont {{Polchinski}}},\ }\href@noop {} {\emph {\bibinfo {title} {{String Theory}}}}\ (\bibinfo {year} {1998})\BibitemShut {NoStop}%
\bibitem [{\citenamefont {{Gibbons}}\ and\ \citenamefont {{Hawking}}(1993)}]{Gibbons:1993}%
  \BibitemOpen
  \bibfield  {author} {\bibinfo {author} {\bibfnamefont {G.~W.}\ \bibnamefont {{Gibbons}}}\ and\ \bibinfo {author} {\bibfnamefont {S.~W.}\ \bibnamefont {{Hawking}}},\ }\href@noop {} {\emph {\bibinfo {title} {{Euclidean quantum gravity}}}}\ (\bibinfo {year} {1993})\BibitemShut {NoStop}%
\bibitem [{\citenamefont {{Smit}}(2002)}]{Smit:2002}%
  \BibitemOpen
  \bibfield  {author} {\bibinfo {author} {\bibfnamefont {J.}~\bibnamefont {{Smit}}},\ }\href@noop {} {\emph {\bibinfo {title} {{Introduction to Quantum Fields on a Lattice}}}}\ (\bibinfo {year} {2002})\BibitemShut {NoStop}%
\bibitem [{\citenamefont {{Turok}}(2014)}]{Turok:2014}%
  \BibitemOpen
  \bibfield  {author} {\bibinfo {author} {\bibfnamefont {N.}~\bibnamefont {{Turok}}},\ }\href {\doibase 10.1088/1367-2630/16/6/063006} {\bibfield  {journal} {\bibinfo  {journal} {New Journal of Physics}\ }\textbf {\bibinfo {volume} {16}},\ \bibinfo {eid} {063006} (\bibinfo {year} {2014})},\ \Eprint {http://arxiv.org/abs/1312.1772} {arXiv:1312.1772 [quant-ph]} \BibitemShut {NoStop}%
\bibitem [{\citenamefont {{Feldbrugge}}\ \emph {et~al.}(2023{\natexlab{a}})\citenamefont {{Feldbrugge}}, \citenamefont {{Jow}},\ and\ \citenamefont {{Pen}}}]{Feldbrugge:2023_Rosen_Morse}%
  \BibitemOpen
  \bibfield  {author} {\bibinfo {author} {\bibfnamefont {J.}~\bibnamefont {{Feldbrugge}}}, \bibinfo {author} {\bibfnamefont {D.~L.}\ \bibnamefont {{Jow}}}, \ and\ \bibinfo {author} {\bibfnamefont {U.-L.}\ \bibnamefont {{Pen}}},\ }\href {\doibase 10.48550/arXiv.2309.12420} {\bibfield  {journal} {\bibinfo  {journal} {arXiv e-prints}\ ,\ \bibinfo {eid} {arXiv:2309.12420}} (\bibinfo {year} {2023}{\natexlab{a}})},\ \Eprint {http://arxiv.org/abs/2309.12420} {arXiv:2309.12420 [quant-ph]} \BibitemShut {NoStop}%
\bibitem [{\citenamefont {{Witten}}(2010)}]{Witten:2010}%
  \BibitemOpen
  \bibfield  {author} {\bibinfo {author} {\bibfnamefont {E.}~\bibnamefont {{Witten}}},\ }\href {\doibase 10.48550/arXiv.1001.2933} {\bibfield  {journal} {\bibinfo  {journal} {arXiv e-prints}\ ,\ \bibinfo {eid} {arXiv:1001.2933}} (\bibinfo {year} {2010})},\ \Eprint {http://arxiv.org/abs/1001.2933} {arXiv:1001.2933 [hep-th]} \BibitemShut {NoStop}%
\bibitem [{\citenamefont {Arnold}\ \emph {et~al.}(2012)\citenamefont {Arnold}, \citenamefont {Gusein-Zade},\ and\ \citenamefont {Varchenko}}]{Arnold:2012}%
  \BibitemOpen
  \bibfield  {author} {\bibinfo {author} {\bibfnamefont {E.}~\bibnamefont {Arnold}}, \bibinfo {author} {\bibfnamefont {S.}~\bibnamefont {Gusein-Zade}}, \ and\ \bibinfo {author} {\bibfnamefont {A.}~\bibnamefont {Varchenko}},\ }\href {https://books.google.co.uk/books?id=ldzAmhSf5wAC} {\emph {\bibinfo {title} {Singularities of Differentiable Maps, Volume 2: Monodromy and Asymptotics of Integrals}}},\ Modern Birkh{\"a}user Classics\ (\bibinfo  {publisher} {Birkh{\"a}user Boston},\ \bibinfo {year} {2012})\BibitemShut {NoStop}%
\bibitem [{\citenamefont {{Feldbrugge}}\ \emph {et~al.}(2023{\natexlab{b}})\citenamefont {{Feldbrugge}}, \citenamefont {{Pen}},\ and\ \citenamefont {{Turok}}}]{Feldbrugge:2023}%
  \BibitemOpen
  \bibfield  {author} {\bibinfo {author} {\bibfnamefont {J.}~\bibnamefont {{Feldbrugge}}}, \bibinfo {author} {\bibfnamefont {U.-L.}\ \bibnamefont {{Pen}}}, \ and\ \bibinfo {author} {\bibfnamefont {N.}~\bibnamefont {{Turok}}},\ }\href {\doibase 10.1016/j.aop.2023.169255} {\bibfield  {journal} {\bibinfo  {journal} {Annals of Physics}\ }\textbf {\bibinfo {volume} {451}},\ \bibinfo {eid} {169255} (\bibinfo {year} {2023}{\natexlab{b}})}\BibitemShut {NoStop}%
\bibitem [{\citenamefont {{Feldbrugge}}\ and\ \citenamefont {{Turok}}(2020)}]{Feldbrugge:2020}%
  \BibitemOpen
  \bibfield  {author} {\bibinfo {author} {\bibfnamefont {J.}~\bibnamefont {{Feldbrugge}}}\ and\ \bibinfo {author} {\bibfnamefont {N.}~\bibnamefont {{Turok}}},\ }\href {\doibase 10.48550/arXiv.2008.01154} {\bibfield  {journal} {\bibinfo  {journal} {arXiv e-prints}\ ,\ \bibinfo {eid} {arXiv:2008.01154}} (\bibinfo {year} {2020})},\ \Eprint {http://arxiv.org/abs/2008.01154} {arXiv:2008.01154 [gr-qc]} \BibitemShut {NoStop}%
\bibitem [{\citenamefont {{Feldbrugge}}(2023)}]{Feldbrugge:2023_Multiplane}%
  \BibitemOpen
  \bibfield  {author} {\bibinfo {author} {\bibfnamefont {J.}~\bibnamefont {{Feldbrugge}}},\ }\href {\doibase 10.1093/mnras/stad349} {\bibfield  {journal} {\bibinfo  {journal} {\mnras}\ }\textbf {\bibinfo {volume} {520}},\ \bibinfo {pages} {2995} (\bibinfo {year} {2023})},\ \Eprint {http://arxiv.org/abs/2010.03089} {arXiv:2010.03089 [astro-ph.CO]} \BibitemShut {NoStop}%
\bibitem [{\citenamefont {{Jow}}\ \emph {et~al.}(2023)\citenamefont {{Jow}}, \citenamefont {{Pen}},\ and\ \citenamefont {{Feldbrugge}}}]{Jow:2023}%
  \BibitemOpen
  \bibfield  {author} {\bibinfo {author} {\bibfnamefont {D.~L.}\ \bibnamefont {{Jow}}}, \bibinfo {author} {\bibfnamefont {U.-L.}\ \bibnamefont {{Pen}}}, \ and\ \bibinfo {author} {\bibfnamefont {J.}~\bibnamefont {{Feldbrugge}}},\ }\href {\doibase 10.1093/mnras/stad2332} {\bibfield  {journal} {\bibinfo  {journal} {\mnras}\ }\textbf {\bibinfo {volume} {525}},\ \bibinfo {pages} {2107} (\bibinfo {year} {2023})},\ \Eprint {http://arxiv.org/abs/2204.12004} {arXiv:2204.12004 [astro-ph.HE]} \BibitemShut {NoStop}%
\bibitem [{\citenamefont {{Cristoforetti}}\ \emph {et~al.}(2012)\citenamefont {{Cristoforetti}}, \citenamefont {{Di Renzo}},\ and\ \citenamefont {{Scorzato}}}]{Cristoforetti:2012}%
  \BibitemOpen
  \bibfield  {author} {\bibinfo {author} {\bibfnamefont {M.}~\bibnamefont {{Cristoforetti}}}, \bibinfo {author} {\bibfnamefont {F.}~\bibnamefont {{Di Renzo}}}, \ and\ \bibinfo {author} {\bibfnamefont {L.}~\bibnamefont {{Scorzato}}},\ }\href {\doibase 10.1103/PhysRevD.86.074506} {\bibfield  {journal} {\bibinfo  {journal} {\prd}\ }\textbf {\bibinfo {volume} {86}},\ \bibinfo {eid} {074506} (\bibinfo {year} {2012})},\ \Eprint {http://arxiv.org/abs/1205.3996} {arXiv:1205.3996 [hep-lat]} \BibitemShut {NoStop}%
\bibitem [{\citenamefont {{Cristoforetti}}\ \emph {et~al.}(2013)\citenamefont {{Cristoforetti}}, \citenamefont {{Di Renzo}}, \citenamefont {{Mukherjee}},\ and\ \citenamefont {{Scorzato}}}]{Cristoforetti:2013}%
  \BibitemOpen
  \bibfield  {author} {\bibinfo {author} {\bibfnamefont {M.}~\bibnamefont {{Cristoforetti}}}, \bibinfo {author} {\bibfnamefont {F.}~\bibnamefont {{Di Renzo}}}, \bibinfo {author} {\bibfnamefont {A.}~\bibnamefont {{Mukherjee}}}, \ and\ \bibinfo {author} {\bibfnamefont {L.}~\bibnamefont {{Scorzato}}},\ }\href {\doibase 10.48550/arXiv.1312.1052} {\bibfield  {journal} {\bibinfo  {journal} {arXiv e-prints}\ ,\ \bibinfo {eid} {arXiv:1312.1052}} (\bibinfo {year} {2013})},\ \Eprint {http://arxiv.org/abs/1312.1052} {arXiv:1312.1052 [hep-lat]} \BibitemShut {NoStop}%
\bibitem [{\citenamefont {{Fujii}}\ \emph {et~al.}(2013)\citenamefont {{Fujii}}, \citenamefont {{Honda}}, \citenamefont {{Kato}}, \citenamefont {{Kikukawa}}, \citenamefont {{Komatsu}},\ and\ \citenamefont {{Sano}}}]{Fujii:2013}%
  \BibitemOpen
  \bibfield  {author} {\bibinfo {author} {\bibfnamefont {H.}~\bibnamefont {{Fujii}}}, \bibinfo {author} {\bibfnamefont {D.}~\bibnamefont {{Honda}}}, \bibinfo {author} {\bibfnamefont {M.}~\bibnamefont {{Kato}}}, \bibinfo {author} {\bibfnamefont {Y.}~\bibnamefont {{Kikukawa}}}, \bibinfo {author} {\bibfnamefont {S.}~\bibnamefont {{Komatsu}}}, \ and\ \bibinfo {author} {\bibfnamefont {T.}~\bibnamefont {{Sano}}},\ }\href {\doibase 10.1007/JHEP10(2013)147} {\bibfield  {journal} {\bibinfo  {journal} {Journal of High Energy Physics}\ }\textbf {\bibinfo {volume} {2013}},\ \bibinfo {eid} {147} (\bibinfo {year} {2013})},\ \Eprint {http://arxiv.org/abs/1309.4371} {arXiv:1309.4371 [hep-lat]} \BibitemShut {NoStop}%
\bibitem [{\citenamefont {{Alexandru}}\ \emph {et~al.}(2016)\citenamefont {{Alexandru}}, \citenamefont {{Basar}}, \citenamefont {{Bedaque}}, \citenamefont {{Ridgway}},\ and\ \citenamefont {{Warrington}}}]{Alexandru:2016}%
  \BibitemOpen
  \bibfield  {author} {\bibinfo {author} {\bibfnamefont {A.}~\bibnamefont {{Alexandru}}}, \bibinfo {author} {\bibfnamefont {G.}~\bibnamefont {{Basar}}}, \bibinfo {author} {\bibfnamefont {P.~F.}\ \bibnamefont {{Bedaque}}}, \bibinfo {author} {\bibfnamefont {G.~W.}\ \bibnamefont {{Ridgway}}}, \ and\ \bibinfo {author} {\bibfnamefont {N.~C.}\ \bibnamefont {{Warrington}}},\ }\href {\doibase 10.1007/JHEP05(2016)053} {\bibfield  {journal} {\bibinfo  {journal} {Journal of High Energy Physics}\ }\textbf {\bibinfo {volume} {2016}},\ \bibinfo {eid} {53} (\bibinfo {year} {2016})}\BibitemShut {NoStop}%
\bibitem [{\citenamefont {{Alexandru}}\ \emph {et~al.}(2017)\citenamefont {{Alexandru}}, \citenamefont {{Ba{\c{s}}ar}}, \citenamefont {{Bedaque}},\ and\ \citenamefont {{Warrington}}}]{Alexandru:2017}%
  \BibitemOpen
  \bibfield  {author} {\bibinfo {author} {\bibfnamefont {A.}~\bibnamefont {{Alexandru}}}, \bibinfo {author} {\bibfnamefont {G.}~\bibnamefont {{Ba{\c{s}}ar}}}, \bibinfo {author} {\bibfnamefont {P.~F.}\ \bibnamefont {{Bedaque}}}, \ and\ \bibinfo {author} {\bibfnamefont {N.~C.}\ \bibnamefont {{Warrington}}},\ }\href {\doibase 10.1103/PhysRevD.96.034513} {\bibfield  {journal} {\bibinfo  {journal} {\prd}\ }\textbf {\bibinfo {volume} {96}},\ \bibinfo {eid} {034513} (\bibinfo {year} {2017})},\ \Eprint {http://arxiv.org/abs/1703.02414} {arXiv:1703.02414 [hep-lat]} \BibitemShut {NoStop}%
\bibitem [{\citenamefont {{Fukuma}}\ and\ \citenamefont {{Umeda}}(2017)}]{Fukuma:2017}%
  \BibitemOpen
  \bibfield  {author} {\bibinfo {author} {\bibfnamefont {M.}~\bibnamefont {{Fukuma}}}\ and\ \bibinfo {author} {\bibfnamefont {N.}~\bibnamefont {{Umeda}}},\ }\href {\doibase 10.1093/ptep/ptx081} {\bibfield  {journal} {\bibinfo  {journal} {Progress of Theoretical and Experimental Physics}\ }\textbf {\bibinfo {volume} {2017}},\ \bibinfo {eid} {073B01} (\bibinfo {year} {2017})},\ \Eprint {http://arxiv.org/abs/1703.00861} {arXiv:1703.00861 [hep-lat]} \BibitemShut {NoStop}%
\bibitem [{\citenamefont {{Fukuma}}\ \emph {et~al.}(2019)\citenamefont {{Fukuma}}, \citenamefont {{Matsumoto}},\ and\ \citenamefont {{Umeda}}}]{Fukuma:2019}%
  \BibitemOpen
  \bibfield  {author} {\bibinfo {author} {\bibfnamefont {M.}~\bibnamefont {{Fukuma}}}, \bibinfo {author} {\bibfnamefont {N.}~\bibnamefont {{Matsumoto}}}, \ and\ \bibinfo {author} {\bibfnamefont {N.}~\bibnamefont {{Umeda}}},\ }\href {\doibase 10.48550/arXiv.1912.13303} {\bibfield  {journal} {\bibinfo  {journal} {arXiv e-prints}\ ,\ \bibinfo {eid} {arXiv:1912.13303}} (\bibinfo {year} {2019})},\ \Eprint {http://arxiv.org/abs/1912.13303} {arXiv:1912.13303 [hep-lat]} \BibitemShut {NoStop}%
\bibitem [{\citenamefont {{Fukuma}}\ and\ \citenamefont {{Matsumoto}}(2021)}]{Fukuma:2021}%
  \BibitemOpen
  \bibfield  {author} {\bibinfo {author} {\bibfnamefont {M.}~\bibnamefont {{Fukuma}}}\ and\ \bibinfo {author} {\bibfnamefont {N.}~\bibnamefont {{Matsumoto}}},\ }\href {\doibase 10.1093/ptep/ptab010} {\bibfield  {journal} {\bibinfo  {journal} {Progress of Theoretical and Experimental Physics}\ }\textbf {\bibinfo {volume} {2021}},\ \bibinfo {eid} {023B08} (\bibinfo {year} {2021})},\ \Eprint {http://arxiv.org/abs/2012.08468} {arXiv:2012.08468 [hep-lat]} \BibitemShut {NoStop}%
\bibitem [{\citenamefont {{Fujisawa}}\ \emph {et~al.}(2022)\citenamefont {{Fujisawa}}, \citenamefont {{Nishimura}}, \citenamefont {{Sakai}},\ and\ \citenamefont {{Yosprakob}}}]{Fujisawa:2022}%
  \BibitemOpen
  \bibfield  {author} {\bibinfo {author} {\bibfnamefont {G.}~\bibnamefont {{Fujisawa}}}, \bibinfo {author} {\bibfnamefont {J.}~\bibnamefont {{Nishimura}}}, \bibinfo {author} {\bibfnamefont {K.}~\bibnamefont {{Sakai}}}, \ and\ \bibinfo {author} {\bibfnamefont {A.}~\bibnamefont {{Yosprakob}}},\ }\href {\doibase 10.1007/JHEP04(2022)179} {\bibfield  {journal} {\bibinfo  {journal} {Journal of High Energy Physics}\ }\textbf {\bibinfo {volume} {2022}},\ \bibinfo {eid} {179} (\bibinfo {year} {2022})},\ \Eprint {http://arxiv.org/abs/2112.10519} {arXiv:2112.10519 [hep-lat]} \BibitemShut {NoStop}%
\bibitem [{\citenamefont {{Nishimura}}\ \emph {et~al.}(2023)\citenamefont {{Nishimura}}, \citenamefont {{Sakai}},\ and\ \citenamefont {{Yosprakob}}}]{Nishimura:2023}%
  \BibitemOpen
  \bibfield  {author} {\bibinfo {author} {\bibfnamefont {J.}~\bibnamefont {{Nishimura}}}, \bibinfo {author} {\bibfnamefont {K.}~\bibnamefont {{Sakai}}}, \ and\ \bibinfo {author} {\bibfnamefont {A.}~\bibnamefont {{Yosprakob}}},\ }\href {\doibase 10.1007/JHEP09(2023)110} {\bibfield  {journal} {\bibinfo  {journal} {Journal of High Energy Physics}\ }\textbf {\bibinfo {volume} {2023}},\ \bibinfo {eid} {110} (\bibinfo {year} {2023})},\ \Eprint {http://arxiv.org/abs/2307.11199} {arXiv:2307.11199 [hep-th]} \BibitemShut {NoStop}%
\bibitem [{\citenamefont {{Nishimura}}\ \emph {et~al.}(2024)\citenamefont {{Nishimura}}, \citenamefont {{Sakai}},\ and\ \citenamefont {{Yosprakob}}}]{Nishimura:2024}%
  \BibitemOpen
  \bibfield  {author} {\bibinfo {author} {\bibfnamefont {J.}~\bibnamefont {{Nishimura}}}, \bibinfo {author} {\bibfnamefont {K.}~\bibnamefont {{Sakai}}}, \ and\ \bibinfo {author} {\bibfnamefont {A.}~\bibnamefont {{Yosprakob}}},\ }\href {\doibase 10.1007/JHEP07(2024)174} {\bibfield  {journal} {\bibinfo  {journal} {Journal of High Energy Physics}\ }\textbf {\bibinfo {volume} {2024}},\ \bibinfo {eid} {174} (\bibinfo {year} {2024})},\ \Eprint {http://arxiv.org/abs/2404.16589} {arXiv:2404.16589 [hep-lat]} \BibitemShut {NoStop}%
\bibitem [{\citenamefont {{Feldbrugge}}\ \emph {et~al.}(2017)\citenamefont {{Feldbrugge}}, \citenamefont {{Lehners}},\ and\ \citenamefont {{Turok}}}]{Feldbrugge:2017}%
  \BibitemOpen
  \bibfield  {author} {\bibinfo {author} {\bibfnamefont {J.}~\bibnamefont {{Feldbrugge}}}, \bibinfo {author} {\bibfnamefont {J.-L.}\ \bibnamefont {{Lehners}}}, \ and\ \bibinfo {author} {\bibfnamefont {N.}~\bibnamefont {{Turok}}},\ }\href {\doibase 10.1103/PhysRevD.95.103508} {\bibfield  {journal} {\bibinfo  {journal} {\prd}\ }\textbf {\bibinfo {volume} {95}},\ \bibinfo {eid} {103508} (\bibinfo {year} {2017})},\ \Eprint {http://arxiv.org/abs/1703.02076} {arXiv:1703.02076 [hep-th]} \BibitemShut {NoStop}%
\bibitem [{\citenamefont {{Chou}}\ and\ \citenamefont {{Nishimura}}(2024)}]{Chou:2024}%
  \BibitemOpen
  \bibfield  {author} {\bibinfo {author} {\bibfnamefont {C.-Y.}\ \bibnamefont {{Chou}}}\ and\ \bibinfo {author} {\bibfnamefont {J.}~\bibnamefont {{Nishimura}}},\ }\href {\doibase 10.48550/arXiv.2407.17724} {\bibfield  {journal} {\bibinfo  {journal} {arXiv e-prints}\ ,\ \bibinfo {eid} {arXiv:2407.17724}} (\bibinfo {year} {2024})},\ \Eprint {http://arxiv.org/abs/2407.17724} {arXiv:2407.17724 [gr-qc]} \BibitemShut {NoStop}%
\bibitem [{\citenamefont {{Feldbrugge}}\ and\ \citenamefont {{Turok}}(2023)}]{Feldbrugge:2023_Existence}%
  \BibitemOpen
  \bibfield  {author} {\bibinfo {author} {\bibfnamefont {J.}~\bibnamefont {{Feldbrugge}}}\ and\ \bibinfo {author} {\bibfnamefont {N.}~\bibnamefont {{Turok}}},\ }\href {\doibase 10.1016/j.aop.2023.169315} {\bibfield  {journal} {\bibinfo  {journal} {Annals of Physics}\ }\textbf {\bibinfo {volume} {454}},\ \bibinfo {eid} {169315} (\bibinfo {year} {2023})},\ \Eprint {http://arxiv.org/abs/2207.12798} {arXiv:2207.12798 [hep-th]} \BibitemShut {NoStop}%
\bibitem [{\citenamefont {Klauder}(2010)}]{Klauder:2010}%
  \BibitemOpen
  \bibfield  {author} {\bibinfo {author} {\bibfnamefont {J.}~\bibnamefont {Klauder}},\ }\href@noop {} {\emph {\bibinfo {title} {A Modern Approach to Functional Integration}}},\ Applied and Numerical Harmonic Analysis\ (\bibinfo  {publisher} {Birkh{\"a}user Boston},\ \bibinfo {year} {2010})\BibitemShut {NoStop}%
\bibitem [{Note1()}]{Note1}%
  \BibitemOpen
  \bibinfo {note} {When an integral diverges absolutely, $\DOTSI \intop \ilimits@ |f(x)|\protect \mathrm {d}x = \infty $, the convergence of the integral $\DOTSI \intop \ilimits@ f(x) \protect \mathrm {d}x$ subtly depends on the regularization of the integral.}\BibitemShut {Stop}%
\bibitem [{\citenamefont {{Klauder}}(2003)}]{Klauder:2003}%
  \BibitemOpen
  \bibfield  {author} {\bibinfo {author} {\bibfnamefont {J.~R.}\ \bibnamefont {{Klauder}}},\ }in\ \href {\doibase 10.1142/9789812795106_0005} {\emph {\bibinfo {booktitle} {A Garden of Quanta: Essays in Honor of Hiroshi Ezawa. Edited by ARAFUNE J ET AL. Published by World Scientific Publishing Co. Pte. Ltd}}},\ \bibinfo {editor} {edited by\ \bibinfo {editor} {\bibfnamefont {J.}~\bibnamefont {{Arafune}}}\ and\ \bibinfo {editor} {\bibnamefont {{et al.}}}}\ (\bibinfo {year} {2003})\ pp.\ \bibinfo {pages} {55--76}\BibitemShut {NoStop}%
\bibitem [{\citenamefont {{Thompson}}(2011)}]{NIST:2011}%
  \BibitemOpen
  \bibfield  {author} {\bibinfo {author} {\bibfnamefont {I.}~\bibnamefont {{Thompson}}},\ }\href {\doibase 10.1080/00107514.2011.582161} {\bibfield  {journal} {\bibinfo  {journal} {Contemporary Physics}\ }\textbf {\bibinfo {volume} {52}},\ \bibinfo {pages} {497} (\bibinfo {year} {2011})}\BibitemShut {NoStop}%
\bibitem [{\citenamefont {{P{\"o}schl}}\ and\ \citenamefont {{Teller}}(1933)}]{Poschl:1933}%
  \BibitemOpen
  \bibfield  {author} {\bibinfo {author} {\bibfnamefont {G.}~\bibnamefont {{P{\"o}schl}}}\ and\ \bibinfo {author} {\bibfnamefont {E.}~\bibnamefont {{Teller}}},\ }\href {\doibase 10.1007/BF01331132} {\bibfield  {journal} {\bibinfo  {journal} {Zeitschrift fur Physik}\ }\textbf {\bibinfo {volume} {83}},\ \bibinfo {pages} {143} (\bibinfo {year} {1933})}\BibitemShut {NoStop}%
\bibitem [{\citenamefont {{Rosen}}\ and\ \citenamefont {{Morse}}(1932)}]{Rosen:1932}%
  \BibitemOpen
  \bibfield  {author} {\bibinfo {author} {\bibfnamefont {N.}~\bibnamefont {{Rosen}}}\ and\ \bibinfo {author} {\bibfnamefont {P.~M.}\ \bibnamefont {{Morse}}},\ }\href {\doibase 10.1103/PhysRev.42.210} {\bibfield  {journal} {\bibinfo  {journal} {Physical Review}\ }\textbf {\bibinfo {volume} {42}},\ \bibinfo {pages} {210} (\bibinfo {year} {1932})}\BibitemShut {NoStop}%
\bibitem [{\citenamefont {{Kleinert}}\ and\ \citenamefont {{Mustapic}}(1992)}]{Kleinert:1992}%
  \BibitemOpen
  \bibfield  {author} {\bibinfo {author} {\bibfnamefont {H.}~\bibnamefont {{Kleinert}}}\ and\ \bibinfo {author} {\bibfnamefont {I.}~\bibnamefont {{Mustapic}}},\ }\href {\doibase 10.1063/1.529800} {\bibfield  {journal} {\bibinfo  {journal} {Journal of Mathematical Physics}\ }\textbf {\bibinfo {volume} {33}},\ \bibinfo {pages} {643} (\bibinfo {year} {1992})}\BibitemShut {NoStop}%
\bibitem [{\citenamefont {Grosche}\ and\ \citenamefont {Steiner}(1998)}]{Grosche:1998}%
  \BibitemOpen
  \bibfield  {author} {\bibinfo {author} {\bibfnamefont {C.}~\bibnamefont {Grosche}}\ and\ \bibinfo {author} {\bibfnamefont {F.}~\bibnamefont {Steiner}},\ }\href@noop {} {\emph {\bibinfo {title} {Handbook of Feynman Path Integrals}}},\ \bibinfo {series} {Handbook of Feynman Path Integrals}\ No.\ \bibinfo {number} {v. 145}\ (\bibinfo  {publisher} {Springer Berlin Heidelberg},\ \bibinfo {year} {1998})\BibitemShut {NoStop}%
\bibitem [{\citenamefont {{Grosche}}\ and\ \citenamefont {{Steiner}}(1993)}]{Grosche:1993}%
  \BibitemOpen
  \bibfield  {author} {\bibinfo {author} {\bibfnamefont {C.}~\bibnamefont {{Grosche}}}\ and\ \bibinfo {author} {\bibfnamefont {F.}~\bibnamefont {{Steiner}}},\ }\href {\doibase 10.48550/arXiv.hep-th/9302053} {\bibfield  {journal} {\bibinfo  {journal} {arXiv e-prints}\ ,\ \bibinfo {eid} {hep-th/9302053}} (\bibinfo {year} {1993})},\ \Eprint {http://arxiv.org/abs/hep-th/9302053} {arXiv:hep-th/9302053 [hep-th]} \BibitemShut {NoStop}%
\bibitem [{\citenamefont {Schulman}(1975)}]{Schulman:1975}%
  \BibitemOpen
  \bibfield  {author} {\bibinfo {author} {\bibfnamefont {L.}~\bibnamefont {Schulman}},\ }in\ \href@noop {} {\emph {\bibinfo {booktitle} {Functional integration and its applications}}}\ (\bibinfo  {publisher} {Oxford Univ. Press London},\ \bibinfo {year} {1975})\BibitemShut {NoStop}%
\bibitem [{\citenamefont {{Fulling}}\ and\ \citenamefont {{G{\"u}nt{\"u}rk}}(2003)}]{Fulling:2003}%
  \BibitemOpen
  \bibfield  {author} {\bibinfo {author} {\bibfnamefont {S.~A.}\ \bibnamefont {{Fulling}}}\ and\ \bibinfo {author} {\bibfnamefont {K.~S.}\ \bibnamefont {{G{\"u}nt{\"u}rk}}},\ }\href {\doibase 10.1119/1.1509415} {\bibfield  {journal} {\bibinfo  {journal} {American Journal of Physics}\ }\textbf {\bibinfo {volume} {71}},\ \bibinfo {pages} {55} (\bibinfo {year} {2003})}\BibitemShut {NoStop}%
\bibitem [{\citenamefont {{Crank}}\ \emph {et~al.}(1947)\citenamefont {{Crank}}, \citenamefont {{Nicolson}},\ and\ \citenamefont {{Hartree}}}]{Crank:1947}%
  \BibitemOpen
  \bibfield  {author} {\bibinfo {author} {\bibfnamefont {J.}~\bibnamefont {{Crank}}}, \bibinfo {author} {\bibfnamefont {P.}~\bibnamefont {{Nicolson}}}, \ and\ \bibinfo {author} {\bibfnamefont {D.~R.}\ \bibnamefont {{Hartree}}},\ }\href {\doibase 10.1017/S0305004100023197} {\bibfield  {journal} {\bibinfo  {journal} {Proceedings of the Cambridge Philosophical Society}\ }\textbf {\bibinfo {volume} {43}},\ \bibinfo {pages} {50} (\bibinfo {year} {1947})}\BibitemShut {NoStop}%
\bibitem [{\citenamefont {{De Raedt}}(1987)}]{De_Raedt:1987}%
  \BibitemOpen
  \bibfield  {author} {\bibinfo {author} {\bibfnamefont {H.}~\bibnamefont {{De Raedt}}},\ }\href {\doibase 10.1016/0167-7977(87)90002-5} {\bibfield  {journal} {\bibinfo  {journal} {Computer Physics Reports}\ }\textbf {\bibinfo {volume} {7}},\ \bibinfo {pages} {1} (\bibinfo {year} {1987})}\BibitemShut {NoStop}%
\bibitem [{\citenamefont {{Wheeler}}(1957)}]{Wheeler:1957}%
  \BibitemOpen
  \bibfield  {author} {\bibinfo {author} {\bibfnamefont {J.~A.}\ \bibnamefont {{Wheeler}}},\ }\href {\doibase 10.1016/0003-4916(57)90050-7} {\bibfield  {journal} {\bibinfo  {journal} {Annals of Physics}\ }\textbf {\bibinfo {volume} {2}},\ \bibinfo {pages} {604} (\bibinfo {year} {1957})}\BibitemShut {NoStop}%
\end{thebibliography}%

\appendix

%%%%%%%%%%%%%%%%%%%%%%%
\section{Stitching method for Gaussian initial state}\label{ap:gauss}
The solution of the Schrödinger equation can be written as the convolution of the initial state with the propagator. For completeness, we here develop the stitching method for the evolution of a Gaussian initial state. However, note that there exist several efficient methods to solve the time dependent Schrödinger equation for arbitrary initial states including for example the Trotter-Suzuki method \cite{De_Raedt:1987} and the second-order Crank-Nicolson method \cite{Crank:1947}.

When considering the evolution of a Gaussian initial state
\begin{align}
    \psi_T(\bm{x}_1) = \int \psi_0(\bm{x}_0) G(\bm{x}_1,\bm{x}_0,T)\mathrm{d}\bm{x}_0\,,
\end{align}
where the initial state assumes the form
\begin{align}
    \psi_0(\bm{x}_0) =\mathcal{N} e^{-\frac{(\bm{x}_0-\bm{\mu})^2}{4 \sigma^2} + i \bm{p} \cdot \bm{x} _0/ \hbar}\,,
\end{align}
with mean initial position $\bm{\mu}$, spread $\sigma$, mean momentum $\bm{p}$, and normalization $\mathcal{N} = \frac{1}{(2 \pi \sigma^2)^{d/4}}$. The discretized version now looks like 
\begin{align}
    \psi_{T,N}(\bm{x}_1)
    = \mathcal{N} e^{-\frac{i a}{2 \hbar} V(\bm{x}_1)}
    \int_{\mathbb{R}^{d N}} e^{-\frac{(\bm{q}_0-\bm{\mu})^2}{4 \sigma^2} + i \bm{p} \cdot \bm{q} _0/ \hbar} e^{ \frac{i m}{2a \hbar}\left[(\bm{q}_0-\bm{q}_1)^2 + \dots + (\bm{q}_{N-1} - \bm{x}_1)^2\right]  - \frac{i a}{\hbar}\left[\frac{1}{2}V(\bm{q}_0) + V(\bm{q}_1) + \dots +V( \bm{q}_{N-1})\right]} \prod_{n=0}^{N-1} \left(c \mathrm{d} \bm{q}_n\right)\,.
\end{align}
We can see this as the iterative series of integrals, with the initial integral 
\begin{align}
        I_1(\bm{q}_1) = c \int  e^{-\frac{(\bm{q}_0-\bm{\mu})^2}{4 \sigma^2} + i \bm{p} \cdot \bm{q} _0/ \hbar+ \frac{i m}{2a \hbar}(\bm{q}_0-\bm{q}_1)^2  -\frac{i a}{2 \hbar}V(\bm{q}_0)} \mathrm{d}\bm{q}_0\,,
\end{align}
and the propagation rule
\begin{align}
    I_{n+1}(\bm{q}_{n+1}) &= c \int I_{n}(\bm{q}_n)e^{\frac{i m}{2 a \hbar}(\bm{q}_n - \bm{q}_{n+1})^2 - \frac{i a}{\hbar} V(\bm{q}_n)} \mathrm{d}\bm{q}_n\,,
\end{align}
yielding 
\begin{align}
    \psi_{T,N}(\bm{x}_1) &= \mathcal{N} e^{-\frac{i a}{2 \hbar} V(\bm{x}_1)} I_N(\bm{x}_1)\,.
\end{align}
The initial state regularizes the integrals. The first integral 
\begin{align}
    I_1(\bm{q}_1) = c \int  e^{-\frac{(\bm{q}_0-\bm{\mu})^2}{4 \sigma^2} + i \bm{p} \cdot \bm{q} _0/ \hbar+ \frac{i m}{2a \hbar}(\bm{q}_0-\bm{q}_1)^2  -\frac{i a}{2 \hbar}V(\bm{q}_0)} \mathrm{d}\bm{q}_0\,
\end{align}
asymptotes to $0$ in the limit $\lVert \bm{q}_1\rVert \to \infty$. We can thus numerically evaluate it on a lattice in $\bm{q}_1$ and evaluate the subsequent integrals using fast Fourier transformations
\begin{align}
    I_{n+1}(\bm{q}_{n+1})=  \mathcal{F}^{-1}\left[ \mathcal{F}\left( I_{n}(\bm{q}_n) e^{ - \frac{i a}{\hbar} V(\bm{q}_n)}\right) e^{-\frac{i a \hbar \bm{k}^2}{2m}} \right]
\end{align}
until we reach $I_N$, to obtain the evolved state.

\end{document}